\documentclass[]{prlab}

\usepackage[toc,page,header]{appendix}
\microtypesetup{expansion=false}
\usepackage{graphicx}
\usepackage{subcaption}
\usepackage{booktabs}
\usepackage{hyperref}
\usepackage{bm}
\usepackage{amsmath}
\usepackage{amssymb}
\usepackage{mathtools}
\usepackage{amsthm}
\usepackage{needspace}
\usepackage{xspace}
\usepackage{wrapfig}
\usepackage{multirow}
\usepackage{tabularx}
\usepackage{colortbl}
\usepackage{pifont}
\usepackage{algorithm}
\usepackage{algorithmic}
\usepackage{float}
\usepackage{listings}
\usepackage[varqu]{inconsolata}
\usepackage{upquote}
\tcbuselibrary{listings,breakable,skins}
\tcbset{promptstyle/.style={listing only, listing options={style=instruction}, enhanced, breakable,
  colback=white, colframe=seedblue, boxrule=0.5pt, arc=2mm,
  left=8pt, right=8pt, top=5pt, bottom=5pt, before skip=6pt, after skip=10pt}}
\newtcbinputlisting{\promptfile}[1]{listing file={#1}, promptstyle}
\newtcblisting{promptbox}{promptstyle}

\graphicspath{{fig/}}
\newcommand{\vibegame}{VibeGame\xspace}
\definecolor{firstbg}{HTML}{DCEFE3}
\definecolor{secondbg}{HTML}{FCEBD3}
\newcommand{\first}[1]{\cellcolor{firstbg}\textbf{#1}}
\newcommand{\second}[1]{\cellcolor{secondbg}\underline{#1}}

\title{\fontsize{16}{20}\selectfont VibeGame: Prompt-to-Game Development with AI-Native\\Engine and Self-Evolving Adversarial Agent Team}

\author[1,*]{Wenbo~Hu}
\author[1,*]{Ken~Li}
\author[1,*]{Jiazhe~Wei}
\author[2]{Yukang~Cao}
\author[1]{Weiyi~Hong}
\author[1]{Jiayi~Dai}
\author[1]{Chenjun~Bai}
\author[1]{Jiajun~Liang}
\author[1]{Yucheng~Liao}
\author[1]{Ruichuan~An}
\author[1]{Zeyu~Lou}
\author[3]{Haofan~Wang}
\author[1]{Yueming~Lyu}
\author{Jingkang~Yang}
\author[2]{Ziwei~Liu}
\author[1,\dagger]{Chenyang~Si}

\affiliation[1]{PRLab, Nanjing University}
\affiliation[2]{S-Lab, Nanyang Technological University}
\affiliation[3]{Liblib.ai}

\contribution[*]{Equal Contribution}
\contribution[\dagger]{Corresponding Author}

\date{\today}
\checkdata[Project Page]{\url{https://vibegame.tettet.org}}
\checkdata[Code]{\url{https://github.com/tettethu/VibeGame}}
\correspondence{\href{mailto:chenyang.si@nju.edu.cn}{chenyang.si@nju.edu.cn}}

\teaser{%
  \centering
  \includegraphics[width=\linewidth]{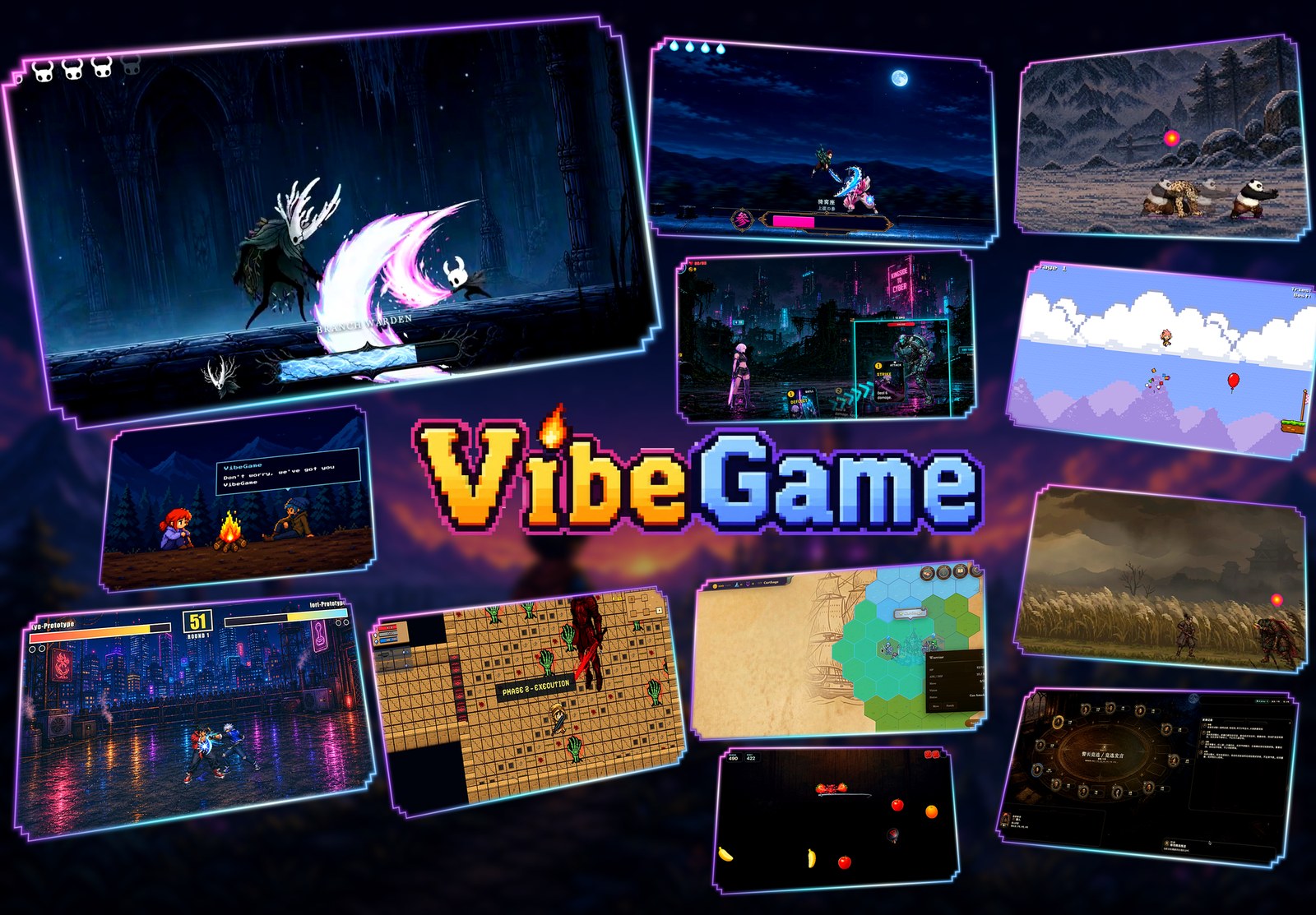}
  \captionof{figure}{Games created and edited by \vibegame from prompts, covering text-only creation, multimodal-input creation, IP transfer, genre transfer, and rule restructuring across diverse 2D game genres.}
  \label{fig:teaser}
}

\abstract{%
Large language model agents have enabled software development through natural-language interaction, but extending this paradigm to end-to-end game development remains challenging.
Building a playable game requires coordinating code, scenes, assets, animations, physics, and configurations, while allowing developers to validate their integrated behavior through interactive testing.
We introduce VibeGame, a framework for creating complete 2D game projects from natural-language prompts and optional reference images.
Its AI-native game engine represents project state as schema-validated structured text and provides frame-synchronous runtime control, enabling agents to inspect game states, execute actions, and verify behavior at their own pace.
Built upon this interface, we design an Adversarial Agent Team of eight specialized agents to separate implementation from review, so that problems are found and fixed through multi-round loops.
For each conducted and accepted project, we further introduce a training-free self-evolution mechanism to distill it into reusable skeletons, modules, and collaboration contracts for future development.
We evaluate prompt-assets-to-game (PA2G), where agents are given an asset library, and prompt-to-game (P2G), where agents generate their own art.
On a GameCraft-Bench subset, VibeGame in P2G consistently outperforms a range of agent harnesses in either P2G or PA2G with the same underlying model, including GPT-6.1-Sol.
}

\begin{document}
\maketitle
\setlength{\parskip}{6pt}

\section{Introduction}
\label{sec:intro}

Games are among the most demanding types of software to develop: they must sustain interactive worlds in real time, maintain a coherent visual identity, and enforce consistent, balanced rules.
Meeting these requirements has traditionally required months of collaboration among designers, programmers, and artists using game engines such as Unity~\citep{unity}, Unreal Engine~\citep{unrealengine}, and Godot~\citep{godot}.
The required time and resources put game development out of reach for many creators.

Artificial intelligence (AI) has begun to lower this barrier from two directions, yet neither fully supports practical game creation.
World models such as Genie~\citep{bruce2024genie} generate interactive experiences directly from video but produce no editable source code, while large language models (LLMs) integrated into commercial engines, as in DreamGarden~\citep{earle2024dreamgarden} for Unreal, face project state scattered across graphical interfaces and binary assets that agents cannot consistently inspect or modify.
Neither produces a complete game project that can be built, played, inspected, versioned, and iteratively revised.

Meanwhile, LLM-based coding agents~\citep{anthropic2025claudecode,openai2025codex} have made ``vibe coding'' increasingly accessible, allowing non-programmers to describe software in natural language and refine it through iterative feedback.
Games are the natural next step, and we formulate this challenge as \textbf{Prompt-to-Game (P2G) Development}: given only a natural language description and optional reference images, the system must produce a complete playable game project, discovering and fixing its problems along the way.

Recent systems and benchmarks have made progress toward this goal.
AutoUE~\citep{autoUE2026} and OpenGame~\citep{opengame2026} build games end to end with agents, and GameDevBench~\citep{chi2026gamedevbench} and GameCraft-Bench~\citep{gamecraftbench2026} evaluate such abilities, the latter also with a provided asset library, a setting we call Prompt-Assets-to-Game (PA2G).
However, most of these efforts reserve runtime checks for final evaluation or limit them to verifying that a game builds and runs.

Yet many game issues emerge only during play, making interactive testing and refinement essential to playability.
Additionally, fixing one project artifact can introduce inconsistencies across interdependent code, scenes, assets, animations, physics, and configurations.

Addressing these challenges reliably and carrying validated solutions forward to future projects requires three tightly coupled capabilities.
First, an \textbf{agent-operable project state} must expose the structure and dependencies among scripts, scenes, assets, animations, physics, and configurations, because direct file access alone cannot prevent a seemingly local edit from violating hidden relationships elsewhere.
Second, \textbf{interaction-grounded verification} is needed because a structurally valid project can still fail at runtime when player input, physics, animation, and game logic interact; since execution unfolds faster than an agent can reason about it, verification requires controllable execution that can be paused, advanced, and examined.
Third, the system must \textbf{preserve validated development experience}: so that reusable project structures, features, and coordination procedures can guide subsequent work, while unverified or project-specific content must remain isolated.

To address these requirements together, we introduce \textbf{VibeGame}, a unified framework that connects project construction, runtime verification, and experience reuse within a development loop, as illustrated in Figure~\ref{fig:overview}.
\textbf{(1)} An \textbf{AI-native game engine} exposes scenes, nodes, assets, scripts, and configurations as structured text with explicit references and schema validation, and its frame-synchronous runtime control lets agents inspect game states, inject actions, and observe their consequences.
\textbf{(2)} Built upon this interface, an \textbf{Adversarial Agent Team (AAT)} separates implementation roles from auditor, player, and reviewer roles, which challenge the game from structural, behavioral, and project-level perspectives; the team iterates review and revision until the project satisfies the intended design and functionality.
\textbf{(3)} Once the project is accepted, a \textbf{training-free self-evolution mechanism} distills its reusable experience into project skeletons, verified modules, and collaboration contracts (Section~\ref{sec:evolving}), which subsequent projects adapt and revalidate rather than reuse blindly.

\begin{figure}[t]
  \centering
  \includegraphics[width=\linewidth]{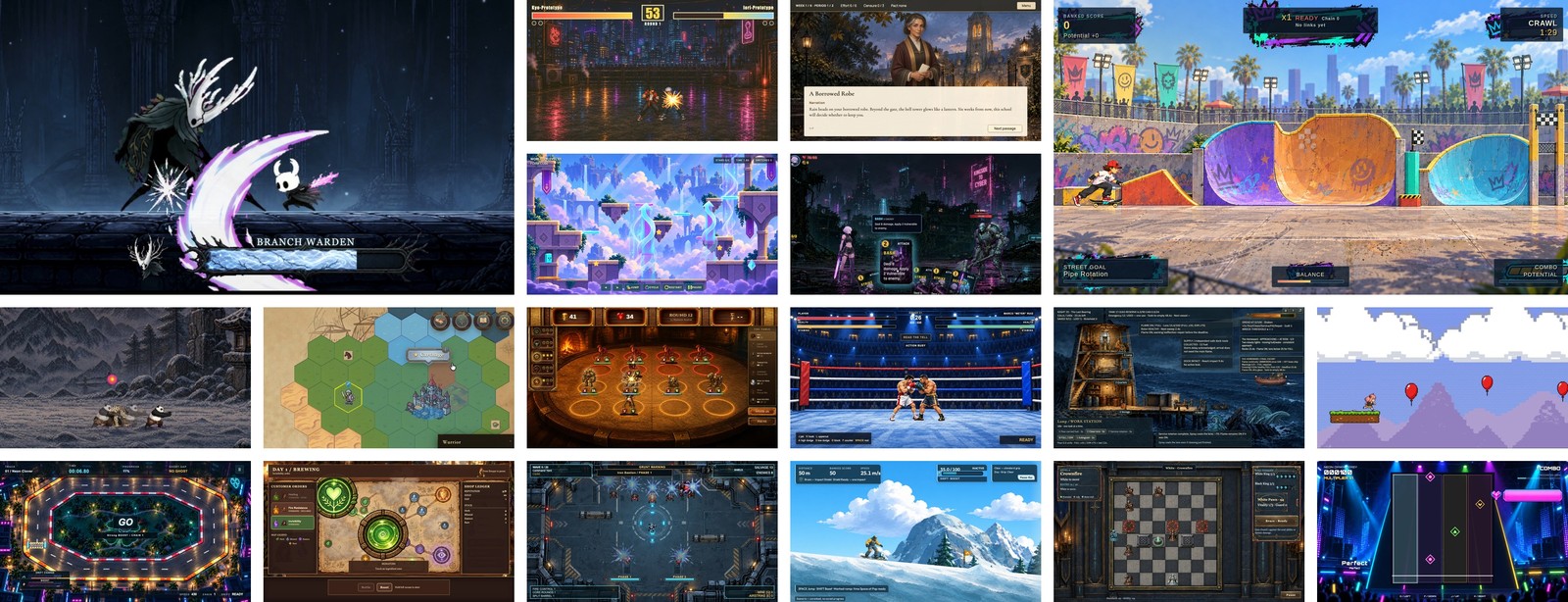}
  \caption{\textbf{Games created by \vibegame from prompts across diverse 2D game genres.}}
  \label{fig:games}
\end{figure}
Figures~\ref{fig:teaser} and~\ref{fig:games} present some examples produced by VibeGame.
And with the same underlying model, \vibegame in P2G outperforms a range of agent harnesses in either P2G or PA2G. Ablation studies further show that removing either the AI-native engine or the adversarial review roles substantially lowers the overall score, and that the experience distilled through self-evolution with a stronger model can be transferred to improve a weaker one.

In summary, our contributions are three-fold:
\begin{itemize}
  \item We introduce \vibegame, a framework for \textbf{prompt-to-game development} that transforms natural-language prompts into complete, playable game projects, enabling rapid game creation and subsequent iterative refinement.
  \item At the core of our method is an \textbf{AI-native game engine} that enables coding agents to directly inspect and modify project state and control runtime execution without relying on graphical editors. Building on this engine, we introduce an \textbf{Adversarial Agent Team} that separates implementation from review, together with a \textbf{training-free self-evolution} mechanism that distills validated development experience into reusable knowledge for future projects.
  \item We comprehensively evaluate \vibegame against other mainstream agent harnesses in both P2G and PA2G settings, demonstrating superior performance over all of them.
\end{itemize}

\section{Related Work}
\label{sec:related}

\noindent\textbf{Agentic software development and vibe coding frameworks.} Coding agents such as Claude Code~\citep{anthropic2025claudecode}, Codex~\citep{openai2025codex}, Cursor~\citep{cursor2024}, OpenHands~\citep{wang2025openhands}, and SWE-agent~\citep{yang2024sweagent}, often with ReAct~\citep{yao2022react}-style reasoning and evaluated on benchmarks such as SWE-bench~\citep{jimenez2024swebench}, have moved LLM-based programming to repository-level development.
Vibe coding frameworks such as Trellis~\citep{mindfold2025trellis} externalize specs, tasks, and memory as persistent repository artifacts, and ChatDev~\citep{qian2023chatdev} and MetaGPT~\citep{hong2023metagpt} introduce role specialization into software development.
\vibegame extends these directions to game projects, where code, art assets, scene configurations, and runtime behavior must be maintained jointly and verified by running the game.

\noindent\textbf{Engine-integrated game generation.} Prior work connects LLMs and generative models to game engines and 3D tools: Sketch2Scene~\citep{xu2024sketch2scene} builds 3D game scenes from sketches; UnrealLLM~\citep{songtang2025unrealllm} and DreamGarden~\citep{earle2024dreamgarden} generate content and environments in Unreal Engine (UE); 3Dify~\citep{hayashi20253dify} uses the Model Context Protocol (MCP)~\citep{anthropic2024mcp} or computer use to call Blender~\citep{blender}, Unity, and UE toolchains, enhanced by retrieval-augmented generation (RAG)~\citep{lewis2020rag}; and \citet{hassan2025automated} generate Unity templates from design documents.
AutoUE~\citep{autoUE2026} chains five agents to generate 3D games in UE and uses play-testing for the final evaluation.
These systems mostly work around the graphical user interface (GUI) workflows of commercial engines, whereas web frameworks like Phaser~\citep{phaser} express projects fully in text.
OpenGame~\citep{opengame2026} builds on Phaser with templates and a debug protocol, generates its own assets from the prompt alone, i.e., the P2G setting, and repairs build and test failures until the game builds and runs.
\vibegame further adds an agent-oriented project representation, runtime control, and visual verification, and lets independent roles verify the running game inside the development loop.

\section{The VibeGame Framework}
\label{sec:overview}

\begin{figure*}[t]
  \centering
  \includegraphics[width=\textwidth]{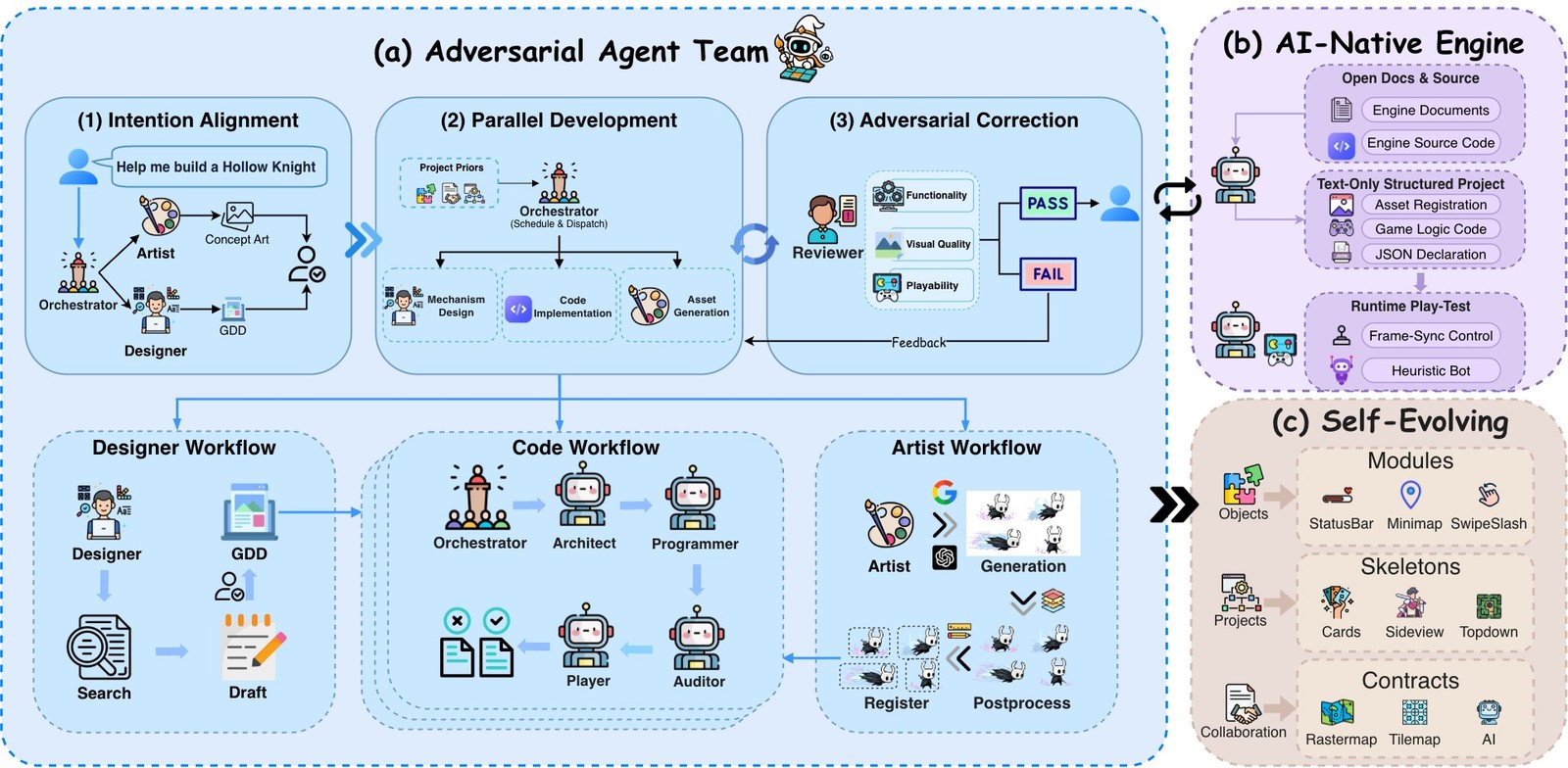}
  \caption{\textbf{System overview of \vibegame.} (a) The Adversarial Agent Team turns a user request into an accepted game through intention alignment, parallel development, and adversarial correction. (b) The AI-native game engine exposes a structured project representation and frame-synchronous runtime control for implementation and play-testing. (c) Training-free self-evolution extracts skeletons, modules, and collaboration contracts from accepted projects and returns them to future development as project priors.}
  \label{fig:overview}
\end{figure*}

VibeGame takes a natural-language request and optional reference images as inputs and produces a complete 2D game project that can be run, inspected, and modified through later requests.

Figure~\ref{fig:overview} presents the project lifecycle.
The AI-native engine provides the shared project representation and runtime interface on which all agents operate (Section~\ref{sec:engine}).
On top of it, the Adversarial Agent Team, eight specialized agents coordinated by an \emph{orchestrator}, turns the request into an accepted game through intention alignment, parallel development, and adversarial correction (Section~\ref{sec:aat}).
After acceptance, the training-free self-evolution mechanism distills the project into skeletons, modules, and collaboration contracts that initialize future runs (Section~\ref{sec:evolving}).

\subsection{AI-Native Game Engine}
\label{sec:engine}

A game engine provides the foundation on which a game project is constructed and executed, but traditional engines are designed for human-operated workflows.
We define a game engine as \textbf{AI-native} when it has three properties that remove the resulting barriers for coding agents.
\textbf{(1) GUI-Independent}: the entire project is represented as plain-text files that agents can inspect and edit directly, rather than engine-specific identifiers or binary files that require graphical editors.
\textbf{(2) Runtime-Accessible}: agents can pause, step through, observe, and intervene in the running game through code rather than GUI-based debugging tools, turning an asynchronous real-time system into a synchronously verifiable process.
\textbf{(3) Source-Available}: agents can inspect the engine's uncompiled source code rather than relying on documentation that may lag behind version changes or omit implementation details.

\textbf{GUI-Independent.}
We build our engine on Phaser because it expresses projects fully in text (Section~\ref{sec:related}).
However, Phaser imposes little structure on how a project is organized, which makes agent-generated projects prone to fragmentation.
We therefore represent each game as a collection of scenes whose configurations are stored in type-checked JavaScript Object Notation (JSON) files.
Each scene is a hierarchy of nodes, the basic building blocks of the project; a node groups its visual appearance, collision geometry, behavior scripts, and configuration into one unit, declared explicitly in JSON fields rather than hidden inside constructors.
Nodes shared across scenes are stored as standalone definition files and instantiated at runtime.
This scene-node-asset hierarchy gives every artifact a clear location and makes cross-file dependencies explicit, so schema validation can detect invalid references, such as missing assets, before execution rather than as runtime failures.

\textbf{Runtime-Accessible.}
A structurally valid project may still fail when player input, physics, animation, and game logic interact during execution.
We therefore expose the running game through a frame-synchronous programmatic interface: agents can pause and resume execution, advance a specified number of frames, inspect or modify runtime properties such as node positions and velocities, inject semantic actions such as ``press jump'', and capture screenshots.
With these controls, agents verify runtime behavior at their own pace through frame-by-frame inspection, scripted test sequences, or heuristic bots that react during real-time play.

\textbf{Source-Available.}
Even with structured project files and runtime control, agents may still encounter engine behavior that the documentation does not fully explain, particularly after an interface change or in uncommon functionality.
We therefore copy the engine's complete, uncompiled source code into the project directory at initialization, so agents can look up the exact application programming interfaces (APIs), assumptions, and implementation details needed for the current task.

\begin{table*}[t]
  \centering
  
  \caption{\textbf{Results on the GameCraft-Bench subset.} DSH denotes DeepSeek Harness. M, D, V, and A denote Core Mechanics, Content Depth, Functional Visuals, and Art \& Presentation, respectively; Overall is the weighted score defined in Appendix~\ref{sec:eval-details}. Bold numbers in green cells and underlined numbers in amber cells denote the best and second-best scores per column within each model, respectively.}
  \begin{tabular*}{\linewidth}{@{\extracolsep{\fill}}lllccccc@{}}
    \toprule
    \textbf{Harness} & \textbf{Model} & \textbf{Input} & \textbf{M} & \textbf{D} & \textbf{V} & \textbf{A} & \textbf{Overall} \\
    \midrule
    OpenGame~\citep{opengame2026}     & GPT-5.6-Luna & P2G  & 36.78 & 25.57 & 23.30 & \second{44.57} & 33.58 \\
    Qwen Code~\citep{qwencode2025}    & GPT-5.6-Luna & PA2G & 46.38 & 32.18 & 27.88 & 37.15 & 35.45 \\
    Kimi Code~\citep{kimicode2026}    & GPT-5.6-Luna & P2G  & 45.26 & 32.13 & 31.00 & 38.34 & 36.14 \\
    Kimi Code & GPT-5.6-Luna & PA2G & 47.54 & 31.86 & 29.67 & 38.46 & 36.22 \\
    Codex~\citep{openai2025codex}     & GPT-5.6-Luna & PA2G & 50.72 & 33.32 & 28.33 & 36.87 & 36.47 \\
    Codex     & GPT-5.6-Luna & P2G  & 45.22 & 34.47 & 31.54 & 37.65 & 36.77 \\
    DSH~\citep{deepseek2026dsh}       & GPT-5.6-Luna & PA2G & \second{52.96} & \second{36.13} & 32.40 & 35.62 & 37.95 \\
    Qwen Code & GPT-5.6-Luna & P2G  & 48.68 & 34.44 & \second{35.51} & 38.72 & \second{38.29} \\
    \vibegame & GPT-5.6-Luna & P2G  & \first{58.31} & \first{43.15} & \first{39.07} & \first{57.82} & \first{49.99} \\
    \midrule
    Codex     & GPT-6.1-Sol  & PA2G & 80.21 & 58.42 & 57.51 & 66.52 & 64.44 \\
    OpenGame  & GPT-6.1-Sol  & P2G  & 81.58 & 58.26 & 57.78 & 67.17 & 64.86 \\
    Qwen Code & GPT-6.1-Sol  & PA2G & 82.42 & 58.06 & 57.50 & 67.97 & 65.15 \\
    Qwen Code & GPT-6.1-Sol  & P2G  & 83.75 & 58.89 & 60.20 & \second{69.90} & 66.71 \\
    Codex     & GPT-6.1-Sol  & P2G  & \first{87.18} & \second{63.05} & \first{62.48} & 69.25 & \second{68.80} \\
    \vibegame & GPT-6.1-Sol  & P2G  & \second{86.26} & \first{67.81} & \second{62.29} & \first{76.97} & \first{72.99} \\
    \bottomrule
  \end{tabular*}
  \label{tab:main}
\end{table*}

\subsection{Adversarial Agent Team}
\label{sec:aat}

The engine provides the interfaces required to construct and test a game, but it does not determine what should be built, how development should be organized, or when the result is satisfactory.
Assigning these responsibilities to a single agent would also require the same role to both produce and evaluate its own work, making behavioral failures and visual defects during gameplay easier to overlook.
VibeGame therefore introduces an \textbf{Adversarial Agent Team} (AAT) comprising eight specialized roles: orchestrator, designer, artist, architect, programmer, auditor, player, and reviewer.
We define each agent's responsibilities through role-specific prompts and execution hooks, and use persistent project artifacts to preserve decisions and communicate them across the team.

As shown in Figure~\ref{fig:overview}(a), the team organizes development into three stages: \textbf{Intention Alignment}, \textbf{Parallel Development}, and \textbf{Adversarial Correction}.
Throughout this process, the orchestrator maintains the development state and coordinates the specialized roles.

\subsubsection{Intention Alignment}
\label{sec:intention-alignment}

Before development begins, the designer draws on game-design principles and genre conventions to expand requests such as ``make a Hollow Knight-style boss fight'' into a complete Game Design Document (GDD) that defines gameplay mechanics, progression rules, and numerical systems.
In parallel, the artist uses image-generation models to create concept art that establishes a consistent visual direction for subsequent assets.
After user confirmation, the GDD and concept art form the shared ground truth for all downstream development and evaluation.
During development, any task that introduces or changes a design decision returns to the designer, who drafts it, confirms it with the user, and merges it into the GDD.

\subsubsection{Parallel Development}
\label{sec:parallel-development}

\textbf{Code Workflow.}
Code is implemented by a pipeline of four agents: architect, programmer, auditor, and player.
The orchestrator first decomposes the game into parallel components, retrieves relevant skeletons, modules, and contracts from the self-evolution repository (Section~\ref{sec:evolving}) as project priors, and prepares a Product Requirements Document (PRD) that defines the scope and acceptance criteria of each component.
The architect then plans the implementation strategy and its verification procedure, and after the orchestrator approves the plan, the programmer implements it.
The auditor checks the result statically with the engine's consistency-checking tools and against the design specification.
Finally, the player runs the game through the engine's runtime interface under frame-synchronous control and uses a vision model to assess whether the rendered result matches the intended appearance.
A sub-task is complete only after it passes both static auditing and runtime play-testing.

\textbf{Artist Workflow.} 
The artist agent converts visual requirements from the orchestrator into engine-ready assets.
It invokes image-generation models, including GPT-Image-2~\citep{openai2026gptimage2} and Nano Banana Pro~\citep{google2025nanobananapro}, to create the initial artwork, then applies post-processing operations such as background removal and size normalization.
It then registers the processed assets with the engine, recording their paths, dimensions, and usage.

\subsubsection{Adversarial Correction}
\label{sec:adversarial-correction}

After all sub-tasks are completed, an independent reviewer acts as the final quality gate and evaluates the complete project on \emph{functionality} (whether the game behaves as specified), \emph{visual quality} (asset completeness, stylistic consistency, and screen clarity), and \emph{playability} (e.g., control responsiveness and the clarity of combat feedback).
If the project fails any criterion, the reviewer rejects it with specific reasons, and the orchestrator converts them into new repair tasks; this build-review cycle continues until the project satisfies all acceptance criteria.
The process runs automatically by default, and the orchestrator incorporates any user feedback into subsequent development.

\subsection{Training-free Self-Evolution}
\label{sec:evolving}

The AI-native game engine and AAT produce accepted game projects that reflect both agent-side implementation and user feedback.
\vibegame distills reusable experience from these validated projects with a training-free self-evolution mechanism and uses it to initialize future development runs, preserving implicit development knowledge that general-purpose agents lose across independent runs.
We organize the experience into three complementary forms: skeletons, modules, and contracts.

\textbf{Skeletons.}
A skeleton is a runnable project template associated with a particular game type.
It removes project-specific art assets but preserves the code architecture, tuning values, visual proportions, scene flow, and runtime structure of the original project.
Each skeleton also includes error notes and an art pack.
The error notes summarize recurring failures and their verified fixes, while the art pack records reusable visual-production recipes, including asset-generation prompts and post-processing conventions for the corresponding game type.

\textbf{Modules.}
A module is a self-contained behavior script that can be attached to any node (Section~\ref{sec:engine}) and implements functionality reusable across game types.
In addition to the underlying behavior, a module may include self-checking logic that converts previously observed runtime failures into static validation rules executed before the game starts.

\textbf{Contracts.} 
A contract is a document that adapts the general AAT workflow to a recurring feature or development pattern.
It specifies the role-specific inputs, outputs, dependencies, and verification criteria required to implement that feature, enabling agents to coordinate through a validated procedure rather than relying only on the generic build-and-review loop.
For example, a map contract defines how the designer specifies map requirements, how the programmer represents the map in project files, how the artist supplies the required assets, and how the auditor and player verify the completed result.

\section{Experiments}
\label{sec:experiments}

\begin{figure*}[t]
  \centering
  \includegraphics[width=\textwidth]{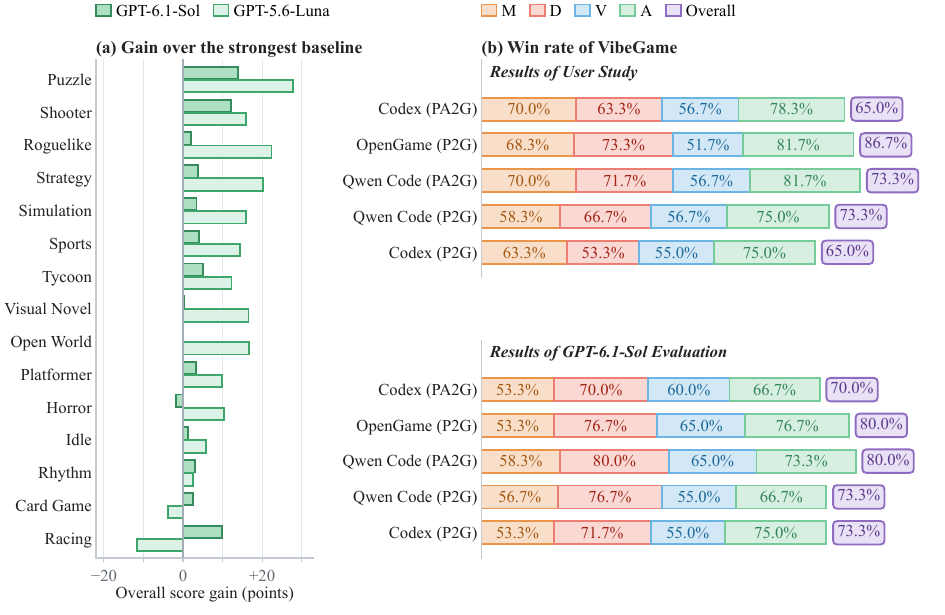}
  \caption{\textbf{Results by game type and pairwise win rates.} (a) Gain of \vibegame over the strongest baseline of each model, Codex (P2G) for GPT-6.1-Sol and Qwen Code (P2G) for GPT-5.6-Luna, on each game type. (b) Win rate of \vibegame against each GPT-6.1-Sol baseline in the user study (top) and under the GPT-6.1-Sol judge (bottom); each bar stacks the win rates in the four rubric categories (M, D, V, A), and the number at its end is the overall win rate.}
  \label{fig:results}
\end{figure*}

\begin{figure*}[t]
  \centering
  \includegraphics[width=\textwidth]{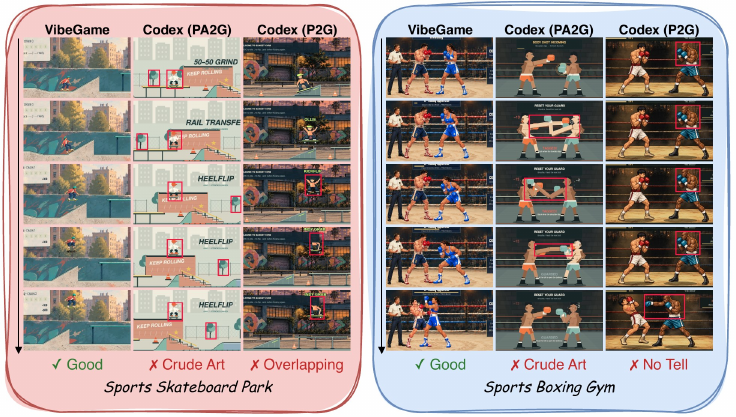}
  \caption{\textbf{Qualitative comparison between \vibegame, Codex (PA2G), and Codex (P2G), all using GPT-6.1-Sol.} Left: Sports Skateboard Park. Right: Sports Boxing Gym. Red boxes mark the defects.}
  \label{fig:qualitative}
\end{figure*}

\subsection{Experiment Setup}
\label{sec:exp-setup}

\paragraph{Benchmark.}
We randomly sample two tasks from each of the 15 game types in GameCraft-Bench~\citep{gamecraftbench2026}, 30 tasks in total, as our \textbf{test set} (Appendix~\ref{sec:task-sets}), and adapt them to Phaser~\citep{phaser}, changing only the engine, the input setting, and a short hint on demo recording, leaving the game descriptions and evaluation rubrics untouched. We then evaluate a wide range of open-source agent harnesses on the test set under the PA2G and P2G settings, giving every harness, including \vibegame, a 24-hour budget per task with its model at the xhigh reasoning effort; in P2G, harnesses without native image generation are given the same image generation model as \vibegame.
Because GPT-5.5~\citep{openai2026gpt55}, the judge used by GameCraft-Bench, makes various judgment errors, we use GPT-6.1-Sol~\citep{openai2026gpt61sol} as the judge instead. Appendix~\ref{sec:exp-details} gives the full experimental details.

\paragraph{Ablation Settings.}
We randomly sample 10 tasks from the test set as the \textbf{ablation set} and compare \vibegame with three variants: (1) \textbf{w/o Engine}, which removes the AI-native game engine and requires the agents to build on Phaser instead; (2) \textbf{w/o AAT}, which removes the adversarial review stage of the agent team, i.e., the auditor, player, and reviewer; and (3) \textbf{w/ Experience}, which adds the experience evolved by \vibegame with GPT-6.1-Sol on 5 tasks outside the ablation set to \vibegame with GPT-5.6-Luna.

\subsection{Results Analysis}
\label{sec:exp-results}

\begin{table}[t]
  \centering
  \caption{\textbf{Ablation on the ablation set (GPT-5.6-Luna).} Columns and marks as in Table~\ref{tab:main}.}
  \setlength{\tabcolsep}{10pt}
  \begin{tabular}{@{}lccccc@{}}
    \toprule
    \textbf{Variant} & \textbf{M} & \textbf{D} & \textbf{V} & \textbf{A} & \textbf{Overall} \\
    \midrule
    \vibegame  & \second{60.73} & \first{48.58} & \first{43.10} & \second{59.26} & \second{53.45} \\
    w/o Engine & 52.16 & 32.95 & 32.98 & 53.22 & 43.12 \\
    w/o AAT    & 57.79 & 41.94 & 34.43 & 58.09 & 48.82 \\
    w/ Experience & \first{68.76} & \second{47.47} & \second{40.42} & \first{66.39} & \first{56.37} \\
    \bottomrule
  \end{tabular}
  \label{tab:ablation}
\end{table}

\paragraph{Main Results.} As shown in Table~\ref{tab:main}, under the same underlying model, general coding agents perform similarly to one another in the PA2G setting, with gaps of at most 2.50 points, and GPT-5.6-Luna~\citep{openai2026gpt56luna} does not turn image generation into a visual advantage in the P2G setting.
OpenGame, although designed for game development, does not outperform general coding agents, suggesting that it may not generalize to tasks beyond its built-in templates or adapt well to the demo-recording requirements of GameCraft-Bench.
In contrast, with GPT-5.6-Luna, \vibegame leads in all four categories and improves the overall score over the strongest baseline by 11.70 points, showing that our harness substantially strengthens weaker models.
With the frontier model GPT-6.1-Sol, \vibegame trails Codex (P2G) by less than one point in Core Mechanics and Functional Visuals but leads it by 4.76 and 7.72 points in Content Depth and Art \& Presentation, outperforming it by 4.19 points overall.
This suggests that frontier models can already implement core mechanics and reasonable visuals on their own, but still fall short in building games with deep content and polished presentation, which is where \vibegame adds its value.
Across game types (Figure~\ref{fig:results}(a)), \vibegame outperforms the strongest baseline of each model on 13 of the 15 types.

\paragraph{User Study.} We further conduct a blind user study with 30 participants of diverse backgrounds, including students from fields such as mathematics and bioinformatics, information technology practitioners, and game enthusiasts. We pair \vibegame with each of the five GPT-6.1-Sol baselines on all 30 tasks, giving 150 pairs, and collect two judgments per pair, 300 in total. In each judgment, a participant reads the game description and the full evaluation rubric of the task, plays the two anonymized games, and picks the better game in each of the four rubric categories and overall. As shown in Figure~\ref{fig:results}(b), human preferences broadly agree with the GPT-6.1-Sol judge, whose pairwise results come from comparing the two games' scores on each task: under both, \vibegame wins against every baseline in every rubric category, with an average overall human win rate of 72.7\%.

\paragraph{Ablation Study.} Table~\ref{tab:ablation} reports the effect of the three coupled components. Removing the AI-native engine or the adversarial review roles of AAT lowers the overall score by 10.33 and 4.63 points, respectively, indicating that both the structured project representation with runtime control and the adversarial review substantially help prompt-to-game development. Adding the engine alone to Codex with GPT-6.1-Sol also raises its overall score by 2.69 points (Appendix~\ref{sec:codex-engine}). Adding the experience evolved by GPT-6.1-Sol to GPT-5.6-Luna raises the overall score by 2.92 points, suggesting that the evolved experience transfers from a stronger model to a weaker one.

\paragraph{Qualitative Analysis.}
We select the qualitative cases from GameCraft-Bench (Figure~\ref{fig:qualitative}).
The left case is Sports Skateboard Park, where the agent needs to implement a trick system in which the player chains flip, grind, and grab tricks into combos.
\vibegame presents aerial tricks such as board flips in a consistent and appealing visual style, while Codex (PA2G) draws the character and environment with programmatic shapes (i.e., simple shapes drawn in code), which gives the game a crude art style.
Codex (P2G) generates a coherent background, but its user interface (UI) overlaps the character, which is also assembled from programmatic shapes.
The right case is Sports Boxing Gym, where the task requires the opponent to telegraph each attack with a visible tell so that the player can read it and respond.
\vibegame fulfills this requirement with complete attack animations and mechanics that include wind-up tells.
Codex (PA2G) also draws its boxers with programmatic shapes and represents each punch by stretching a rectangular arm, which looks abstract rather than like a punching animation.
Codex (P2G) also produces polished visuals but does not show the opponent's wind-up.

\section{Conclusion}
\label{sec:conclusion}

We present \vibegame, a framework for end-to-end prompt-to-game development that integrates an AI-native game engine, an adversarial agent team, and training-free self-evolution.
Under the same underlying model, \vibegame, without any provided assets, outperforms all evaluated agent harnesses; even with the frontier model GPT-6.1-Sol, it improves over the strongest baseline, Codex (P2G), by 4.19 points.

\section*{Limitations}
\label{sec:limitations}

While \vibegame delivers strong results across diverse game genres, we recognize that several directions remain open for future work.
First, we focus on 2D games and have not yet validated the framework in more complex 3D scenes or used it to develop complete commercial games that sustain long-term play.
Second, our AI-native design principles are implemented on a web-based 2D engine; bringing them to established engines such as Godot~\citep{godot}, so that agents can also operate existing game projects, is a promising extension.

\clearpage
\begingroup
\raggedright
\bibliographystyle{unsrtnat}
\bibliography{references}
\endgroup

\clearpage
\beginappendix
\section{Evaluation Details}
\label{sec:eval-details}

We port the evaluation protocol of GameCraft-Bench from Godot to Phaser: each harness submits its game together with replayable input traces; an automated pipeline checks that the game builds, then replays and records every trace, and the judge model scores each recording against the rubric; each requirement's score is then aggregated over the demos as specified by the rubric.
We average the scores of three judge passes to reduce judge variance.

GameCraft-Bench scores each game with a task-specific rubric whose requirements fall into four categories: Core Mechanics ($M$), Content Depth ($D$), Functional Visuals ($V$), and Art \& Presentation ($A$). Each category score is the mean of its requirement scores, and the overall score is
\begin{equation*}
  \mathrm{Overall} = \mathrm{BUILD} \times (w_M M + w_D D + w_V V + w_A A),
\end{equation*}
where $\mathrm{BUILD}$ is 0 if the game cannot be built or played and 1 otherwise.
The weights $(w_M, w_D, w_V, w_A)$ are specified by each task's rubric in GameCraft-Bench.

\begin{table}[H]
  \centering
  \caption{Results on the GameCraft-Bench subset scored by the GPT-5.5 judge (mean of its passes). Rows, columns, and marks as in Table~\ref{tab:main}.}
  \begin{tabular*}{\linewidth}{@{\extracolsep{\fill}}lllccccc@{}}
    \toprule
    \textbf{Harness} & \textbf{Model} & \textbf{Input} & \textbf{M} & \textbf{D} & \textbf{V} & \textbf{A} & \textbf{Overall} \\
    \midrule
    OpenGame  & GPT-5.6-Luna & P2G  & 46.04 & 33.90 & 33.17 & 47.22 & 40.29 \\
    Codex     & GPT-5.6-Luna & P2G  & 58.51 & 45.23 & 48.69 & 47.24 & 48.46 \\
    Kimi Code & GPT-5.6-Luna & P2G  & 61.04 & 43.42 & 52.14 & 47.67 & 48.89 \\
    Qwen Code & GPT-5.6-Luna & PA2G & 66.67 & 46.47 & 51.61 & 49.93 & 51.51 \\
    Kimi Code & GPT-5.6-Luna & PA2G & 65.79 & 44.68 & 50.50 & \second{52.73} & 51.57 \\
    Qwen Code & GPT-5.6-Luna & P2G  & 64.56 & 47.71 & \second{53.30} & 50.17 & 51.97 \\
    Codex     & GPT-5.6-Luna & PA2G & \second{68.51} & 48.49 & 50.51 & 51.27 & 52.81 \\
    DSH       & GPT-5.6-Luna & PA2G & \first{70.05} & \second{48.80} & \first{55.47} & 49.29 & \second{53.19} \\
    \vibegame & GPT-5.6-Luna & P2G  & 64.98 & \first{51.89} & 47.76 & \first{58.72} & \first{55.65} \\
    \midrule
    OpenGame  & GPT-6.1-Sol  & P2G  & 86.57 & 67.68 & 72.44 & 74.36 & 73.60 \\
    Qwen Code & GPT-6.1-Sol  & P2G  & 86.51 & 70.58 & 74.35 & 76.26 & 75.56 \\
    Qwen Code & GPT-6.1-Sol  & PA2G & 88.58 & 70.43 & \first{75.41} & 76.44 & 76.03 \\
    Codex     & GPT-6.1-Sol  & PA2G & 89.57 & 70.57 & 75.12 & 77.29 & 76.49 \\
    Codex     & GPT-6.1-Sol  & P2G  & \first{90.29} & \second{71.73} & \second{75.38} & \second{78.43} & \second{77.44} \\
    \vibegame & GPT-6.1-Sol  & P2G  & \second{90.04} & \first{75.43} & 74.41 & \first{78.51} & \first{78.57} \\
    \bottomrule
  \end{tabular*}
  \label{tab:gpt55-judge}
\end{table}

\subsection{Judge Error Cases (GPT-5.5)}
\label{sec:judge-errors}
\begin{figure}[H]
  \centering
    \includegraphics[width=\linewidth]{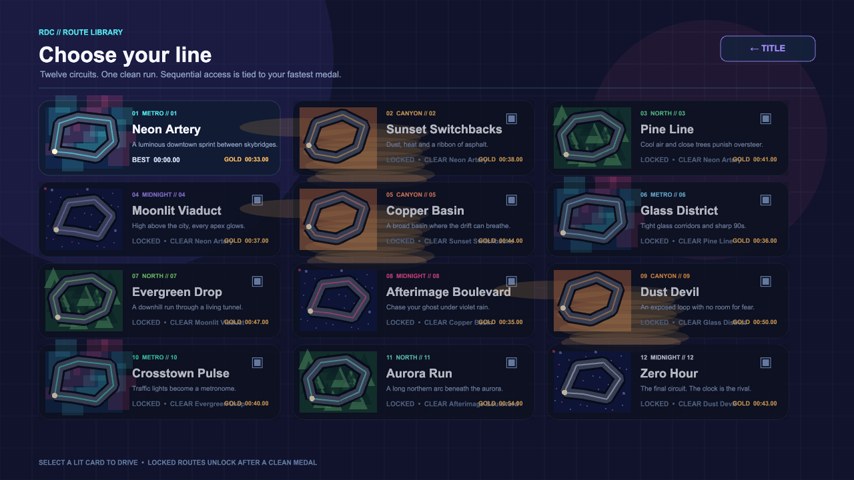}
  \\[4pt]
  \begin{minipage}{\linewidth}\small\raggedright
    \textbf{Requirement (D1).} ``At least 10 distinct tracks with different layouts (varying corner types, widths, and lengths). Tracks are selectable from a track-select screen showing preview thumbnails and medal status. Score 0 if there are fewer than 10 tracks, if tracks share the same layout, or if there is no track selection screen. Score 1 requires content variants to be visually distinct with real authored art (different sprites, illustrated elements). Score 0.5 at most if variants exist but are represented by programmatic shapes, color swaps, or text labels without real asset differences.''\\[4pt]
    \textbf{GPT-5.5} (0.9 / 1.0 / 0.9): ``The route library shows twelve selectable tracks with distinct preview layouts, locked/completed state, and medal/best-time information.''\\[4pt]
    \textbf{GPT-6.1-Sol} (0.0 / 0.0 / 0.0): ``The selection screen lists twelve tracks, but its thumbnails repeatedly reuse four layouts rather than demonstrate at least ten distinct circuits.''
  \end{minipage}
  \caption{Judge error case 1: Kimi Code + GPT-5.6-Luna (P2G) on Racing Drift Circuit. The judged track-select frame: its twelve thumbnails reuse four layouts, one per biome (e.g., the three Canyon tracks share one layout).}
  \label{fig:judge-case1}
\end{figure}

\begin{figure}[H]
  \centering
    \includegraphics[width=\linewidth]{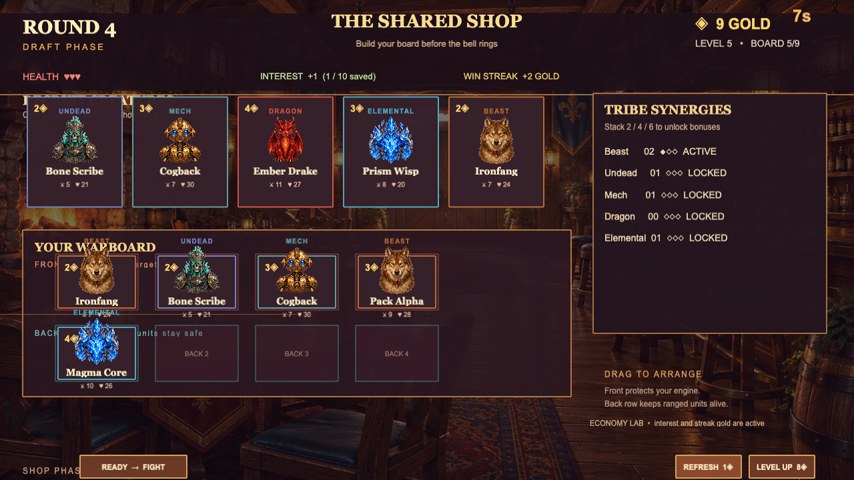}
  \\[4pt]
  \begin{minipage}{\linewidth}\small\raggedright
    \textbf{Requirement (V1).} ``The shop phase clearly shows: available creatures with stats and tribe, the player's board and bench, gold amount, level, and round number. All elements are readable and non-overlapping at 1280x720. Tribal affiliations are visually indicated on creature cards. Score 0 if any core element is missing or unreadable. Score 1 requires the information to be presented through themed, styled UI with real assets. Score 0.5 at most if readability relies on plain text, unstyled labels, or raw geometric shapes.''\\[4pt]
    \textbf{GPT-5.5} (1.0 / 1.0 / 1.0): ``The shop screen clearly shows available creatures with stats and tribes, board and bench areas, gold, level, round number, timer, and themed readable UI at 1280x720.''\\[4pt]
    \textbf{GPT-6.1-Sol} (0.0 / 0.0 / 0.0): ``The themed shop shows creature stats, tribes, gold, level, and round, but no bench is visible and board headings overlap creature labels and cards.''
  \end{minipage}
  \caption{Judge error case 2: OpenGame + GPT-5.6-Luna on Cardgame Autobattler. The board has a front and a back row (BACK 2--4) and no bench; the board heading and row labels are covered by cards.}
  \label{fig:judge-case2}
\end{figure}

\begin{figure}[H]
  \centering
    \includegraphics[width=\linewidth]{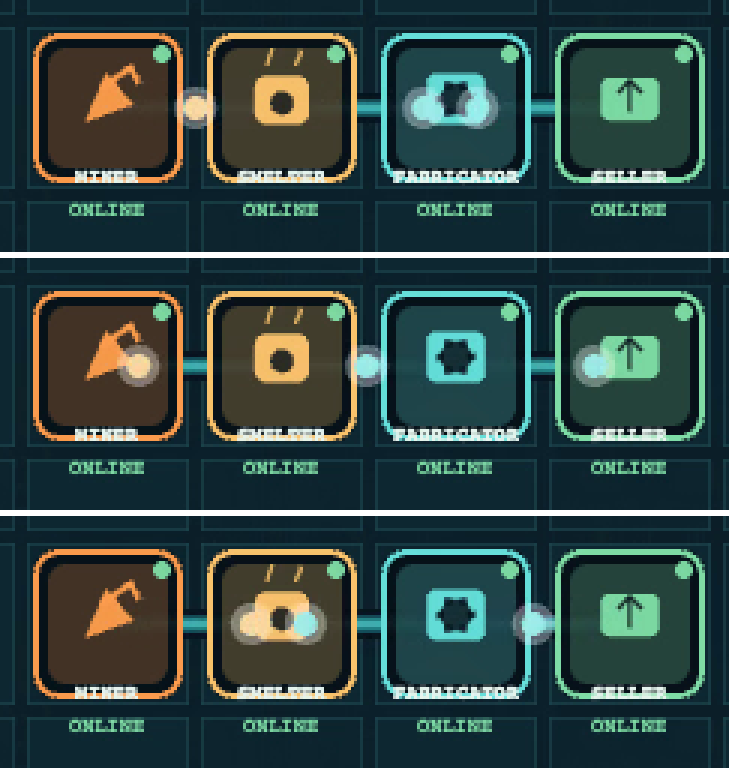}
  \\[4pt]
  \begin{minipage}{\linewidth}\small\raggedright
    \textbf{Requirement (M2).} ``Production chains exist: output from one machine feeds input of the next via visible conveyor belts. Machines must be connected in sequence (miner to smelter to fabricator to seller). Resources visibly flow along conveyors between machines. Score 0 if there are no production chains, if machines work independently without connections, or if resource flow is not visible.''\\[4pt]
    \textbf{GPT-5.5} (0.0 / 0.0 / 0.0): ``Machines are placed near each other, but visible conveyor belts and resources flowing between miner, smelter, fabricator, and seller are not clearly shown.''\\[4pt]
    \textbf{GPT-6.1-Sol} (1.0 / 1.0 / 1.0): ``Two sequential miner-to-smelter-to-fabricator-to-seller chains have visible connecting conveyors with resource particles changing position over time.''
  \end{minipage}
  \caption{Judge error case 3: Kimi Code + GPT-5.6-Luna (P2G) on Idle Factory Planet. Three consecutive judged frames, cropped to one production chain and shown at the source recording's 1280$\times$720 resolution (the judges received them downscaled to 854$\times$480): conveyors connect the four machines and the resource particles move along them.}
  \label{fig:judge-case3}
\end{figure}

We tried both GPT-5.5 and GPT-6.1-Sol as the judge model (both with medium reasoning effort) and found that switching the judge changes both the absolute scores and the ranking (Table~\ref{tab:main} vs.\ Table~\ref{tab:gpt55-judge}).
\vibegame achieves the best overall score under both judges.
Nevertheless, comparing the two judges, we find that GPT-5.5 has severe judgment flaws as a judge model; Figures~\ref{fig:judge-case1}, \ref{fig:judge-case2}, and~\ref{fig:judge-case3} show three examples.

\subsection{Results under GPT-5.5 Judge}
\label{sec:judge-comparison}

Table~\ref{tab:gpt55-judge} reports the results with GPT-5.5 as the judge model.

\section{Benchmark Tasks}
\label{sec:task-sets}

Our test set consists of 30 tasks from GameCraft-Bench~\citep{gamecraftbench2026}, two from each of its 15 game types, as listed in Table~\ref{tab:task-sets}. The ablation set is a 10-task subset of the test set.

\begin{table}[H]
  \centering
  \caption{The 30 benchmark tasks, two per game type. $^\dagger$ marks the 10 tasks of the ablation set.}
  \begin{tabular}{@{}l>{\raggedright\arraybackslash}p{0.64\columnwidth}@{}}
    \toprule
    \textbf{Game type} & \textbf{Tasks} \\
    \midrule
    Card Game    & Autobattler$^\dagger$, Poker Roguelike$^\dagger$ \\
    Horror       & Lighthouse$^\dagger$, Tape Archive \\
    Idle         & Dungeon Guild$^\dagger$, Factory Planet \\
    Open World   & Beast Tamer, Sky Islands \\
    Platformer   & Shape Shift, Thunder Valkyrie \\
    Puzzle       & Gravity Shift$^\dagger$, Pipe Crisis$^\dagger$ \\
    Racing       & Drift Circuit, Trick Runner \\
    Rhythm       & Conductor, Note Highway \\
    Roguelike    & Breach Tactics$^\dagger$, Potion Craft \\
    Shooter      & Mech Sortie, Wave Commander \\
    Simulation   & Air Control, Transit Web \\
    Sports       & Boxing Gym, Skateboard Park$^\dagger$ \\
    Strategy     & Chess Variant, Robot Factory \\
    Tycoon       & Tiny Factory Foreman$^\dagger$, Trading Caravan Tycoon \\
    Visual Novel & Arcane Academy$^\dagger$, Courtroom Clue Trial \\
    \bottomrule
  \end{tabular}
  \label{tab:task-sets}
\end{table}

\section{Experiment Details}
\label{sec:exp-details}

\subsection{Setup}
Every harness, including \vibegame, has a 24-hour budget per task and runs its model at the xhigh reasoning effort. In the P2G setting, harnesses without a native image generation tool are given GPT-Image-2~\citep{openai2026gptimage2}, the same image generation model used by \vibegame, as an external tool together with its documentation.
The judge GPT-6.1-Sol runs with medium reasoning effort, and scoring follows the original benchmark (Appendix~\ref{sec:eval-details}). All results are obtained from our own runs and evaluation rather than reused from GameCraft-Bench, because we changed the engine and modified the task instructions.

For w/ Experience, we let \vibegame with GPT-6.1-Sol build and self-evolve on 5 tasks randomly sampled from outside the ablation set, using the task rubrics and the vision-language model (VLM) judge scores in place of user feedback. We then add the evolved artifacts to the initial project of \vibegame with GPT-5.6-Luna and let it build the ablation tasks.

\subsection{Task Instructions}
\label{sec:instructions}

\lstdefinestyle{instruction}{basicstyle=\ttfamily\scriptsize, breaklines=true, breakatwhitespace=true,
  columns=fullflexible, keepspaces=true, breakindent=0pt, aboveskip=0pt, belowskip=0pt,
  literate={—}{{\textemdash}}1 {–}{{\textendash}}1 {×}{{$\times$}}1 {→}{{$\rightarrow$}}1}

To close the gap between the quality of a game and what its demos show, we add the following demo hint to the original instructions:
\begin{promptbox}
> All scoring is done from these recordings alone — your source code is never read. A feature no demo shows does not count, however well it is built. Use the full 20 s each demo allows rather than cutting off after a few seconds, keep each state on screen long enough to read, and show as many distinct assets and element types as each scene plausibly supports.
\end{promptbox}
We also use Phaser instead of Godot as the game engine, so we do not claim that our results are comparable with those reported on the original GameCraft-Bench.

\subsubsection{\vibegame}
\promptfile{instructions/cardgame-autobattler_vibegame.md}
The first line of the instruction invokes the one-shot mode of \vibegame through the \texttt{vibegame-build} skill, whose prompt is as follows:
\promptfile{instructions/vibegame-build_SKILL.md}

\subsubsection{Baselines, PA2G}
\promptfile{instructions/cardgame-autobattler_pa2g.md}

\subsubsection{Baselines, P2G (Qwen Code, Kimi Code, DSH)}
\promptfile{instructions/cardgame-autobattler_p2g.md}

\subsubsection{Baselines, P2G (Codex)}
\promptfile{instructions/cardgame-autobattler_p2g-codex.md}

\section{Demonstrations}
\label{sec:demo}

We demonstrate \vibegame in two scenarios: creating a game from a new request (\textbf{Game Creation}), which corresponds to the P2G task defined in Section~\ref{sec:intro}, and continuing to edit an existing project (\textbf{Game Edit}).
Unless otherwise noted, the demonstrations use the automatic-by-default setting described in Section~\ref{sec:adversarial-correction}: after the user approves the request, the orchestrator continues development until reviewer acceptance, while the user may still interrupt the process with feedback. 
The prompt boxes in the figures report the final user-approved request rather than interface commands or intermediate clarification turns.

\begin{figure}[H]
  \centering
  \includegraphics[width=\textwidth]{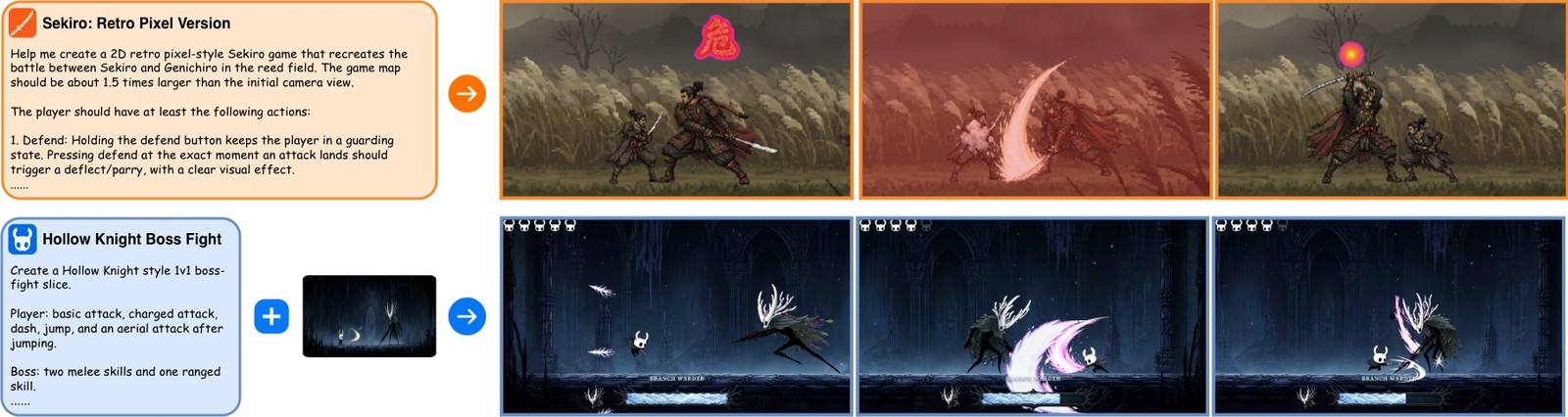}
  \caption{\textbf{Game creation demonstrations.} Top: Pixel Sekiro, a pixel-art boss fight created from a text prompt. Bottom: Hollow Knight boss slice, a side-scrolling boss fight created from a gameplay prompt and a concept image.}
  \label{fig:game-creation}
\end{figure}

\begin{figure}[H]
  \centering
  \includegraphics[width=\textwidth]{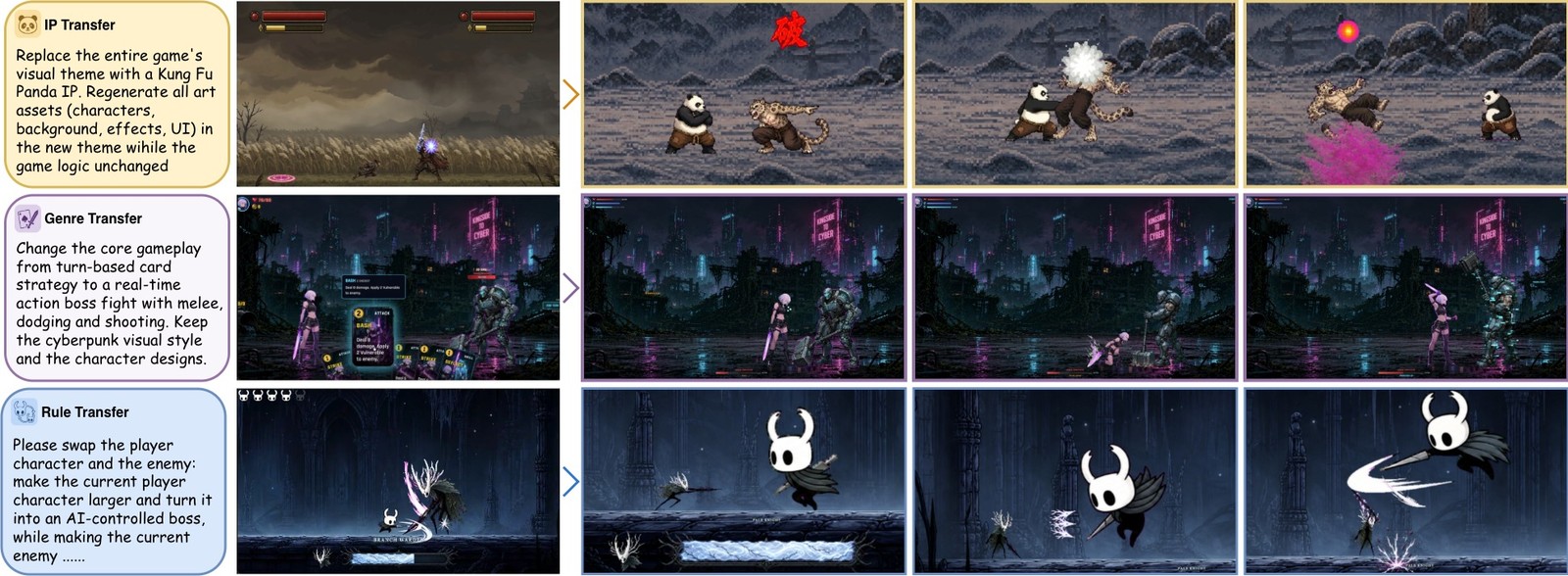}
  \caption{\textbf{Game edit demonstrations.} Top: intellectual property (IP) transfer, replacing the visual theme while preserving gameplay logic. Middle: genre transfer, converting a turn-based card game into a real-time boss fight while preserving its visual identity. Bottom: rule transfer, swapping the player and boss roles while preserving their abilities and animations.}
  \label{fig:game-edit-demo}
\end{figure}

\subsection{Game Creation}
\label{sec:game-creation}

Game creation starts from a text request and optional reference images and exercises the full loop from intention alignment to project delivery. In this paper, we report results for both text-only and multimodal requests.

\subsubsection{Text Input}
\label{sec:text-input}

To test creation driven by a compact text request, we asked the system to build a Sekiro-style 2D boss fight: a retro pixel-art duel recreating the reed-field battle between Sekiro and Genichiro.
The request specified the combat rules in a few sentences: the player can defend, perform frame-perfect deflections, attack, step-dodge, jump, slash in the air, and execute a deathblow once the boss posture is broken; the boss can defend and deflect, and special attacks are made readable through delayed wind-up frames.

The system started from an existing side-scrolling action skeleton, inheriting its scene structure and UI modules. 
However, Sekiro's core is a deflection-posture system entirely different from the skeleton's simple hit-point-based combat, requiring the system to build the combat layer from scratch: both sides have independent posture gauges, defense distinguishes regular blocking from frame-perfect deflection, boss attacks are divided into blockable normal attacks and perilous attacks requiring specific responses, and a full posture gauge triggers a multi-stage deathblow animation.
The final output comprised 12 scripts, 2 character asset packs, 11 visual effect entities, and 1 UI asset pack. 
This case demonstrates how a short request that names a genre convention is expanded into a complete, correctly tuned combat system. 
The first row of Figure~\ref{fig:game-creation} shows the request and resulting combat sequence.

\subsubsection{Multimodal Input}
\label{sec:multimodal-input}

To test multimodal creation, we provided a text request together with an author-created concept image for a Hollow Knight-style 1v1 boss fight. 
The text request specified the mechanical scope: the player has a basic attack, a charged attack, a dash, a jump, and an aerial attack, while the boss combines two melee skills with a ranged attack on a side-scrolling arena 1.5x the camera width. 
The concept image anchored the character designs, the ruined-cathedral arena, and the overall visual atmosphere, which the generated assets were required to match.

The second row of Figure~\ref{fig:game-creation} showcases the multimodal input and resulting game.
We can observe that the delivered slice realizes both sides of this specification. 
The boss gates its melee sweep, ground slam, and arcing projectile on the player's distance, keeps its preferred range, and can be staggered out of idle; the player's five actions cover the full movement-attack loop, including dash invincibility frames and aerial slashes; hit feedback combines screen shake, flash, and spark particles, with mask-style health icons and a full-width boss bar completing the heads-up display (HUD). 
All character and arena art was generated to match the concept image, preserving its silhouette language and cold palette.
This case demonstrates how text can specify gameplay rules while the concept image anchors visual identity.

\subsection{Game Edit}
The next cases begin from accepted projects and alter different layers of the existing game.
Specifically, in this paper, we report three types of editing:
(1) intellectual property (IP) transfer that changes visual assets while preserving mechanics;
(2) genre transfer that replaces the central gameplay while preserving visual identity;
and (3) rule transfer that changes the roles of existing characters while reusing their abilities and animations.

\subsubsection{IP Transfer}
\label{sec:ip-transfer}

Starting from a completed side-scrolling boss fight project, we ask the system to replace the entire game's visual theme with a Kung Fu Panda IP.
The system regenerates all art assets (including characters, effects, UI, and background), and then adjusts configurations for the new sprite size and frame counts. The delivered project retains the original combat logic, illustrating an edit concentrated on the asset and configuration layers. See top row of Figure~\ref{fig:game-edit-demo}. 
This validates the code-asset decoupling in the engine: when the project structure is well designed, IP transfer is accomplished primarily through asset replacement, with no need to rewrite core logic.

\subsubsection{Genre Transfer}
\label{sec:genre-transfer}

Starting from a completed turn-based card game project, we ask the system to change the core gameplay from card strategy to a real-time action boss fight.
This is a deep transformation: the turn-based card-playing mechanic was completely replaced with real-time melee, dodging, and shooting, while the system preserved the original project's cyberpunk visual style and character designs. 
Unlike IP transfer where only assets change, genre transfer requires the system to judge what can be preserved and what must be rewritten, demonstrating the system's ability to perform selective deep modification while maintaining overall consistency (Figure~\ref{fig:game-edit-demo}, middle row).

\subsubsection{Rule Transfer}
\label{sec:rule-transfer}

Starting from the Hollow Knight-style boss fight created from multimodal input in Section~\ref{sec:multimodal-input}, we tested the system's ability to handle unconventional creative requests by asking it to swap the player and boss roles, requiring only that each character keep its existing abilities and animations.
In that source game, the player controlled a small horned warrior, while the antlered boss moved autonomously with a melee slam, a lunge, and a ranged thorn volley.
The edit enlarges the former player character into an AI-driven boss and shrinks the former boss into the player-controlled character, so the player now wields the former boss's moveset while the former player character fights autonomously with its own dash, charged slash, and aerial attack. 
This request is not a simple asset exchange: the system must convert the former boss AI into a player control scheme, turn the former player's manual controls into autonomous boss AI, rebind inputs, and remap the HUD and win/lose flow, all without adding any new art or moves.
Our system completed this transformation (Figure~\ref{fig:game-edit-demo}, bottom row), demonstrating that it can restructure gameplay roles in response to non-standard edit requests.

\section{More Demonstrations}
\label{app:more-demos}

Beyond the cases analyzed in Appendix~\ref{sec:demo}, we used \vibegame to build a broader range of games spanning multiple genres and input modalities. Each game was produced from a natural language prompt, some accompanied by a reference image, exercising the same pipeline described in the main text. Together they illustrate the breadth of game types the system handles, from real-time action and arcade games to turn-based strategy and an LLM-driven social-deduction game.

\subsection*{Fruit Ninja}
\begin{figure}[H]
  \centering
  \includegraphics[width=\linewidth]{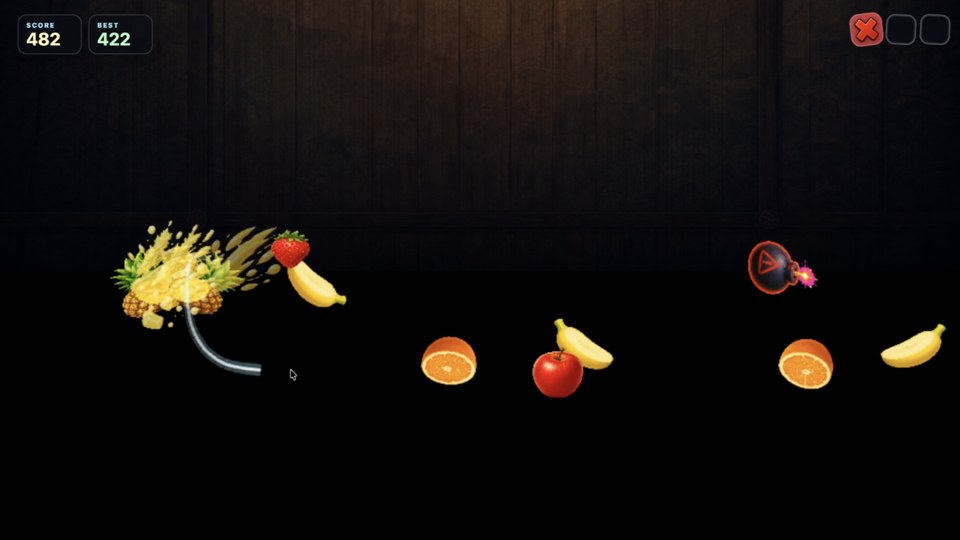}
  \caption{A three-fruit combo sliced mid-air with split halves and juice effects.}
  \label{fig:app-fruit}
\end{figure}

A fruit-slicing arcade game built from a text-only request (Figure~\ref{fig:app-fruit}). Each fruit splits into two physically simulated halves with fruit-specific weight, bounce, and juice effects; the game includes pointer-driven slicing, combo scoring, missed-fruit indicators, and a game-over state.

\subsection*{The King of Fighters}
\begin{figure}[H]
  \centering
  \includegraphics[width=\linewidth]{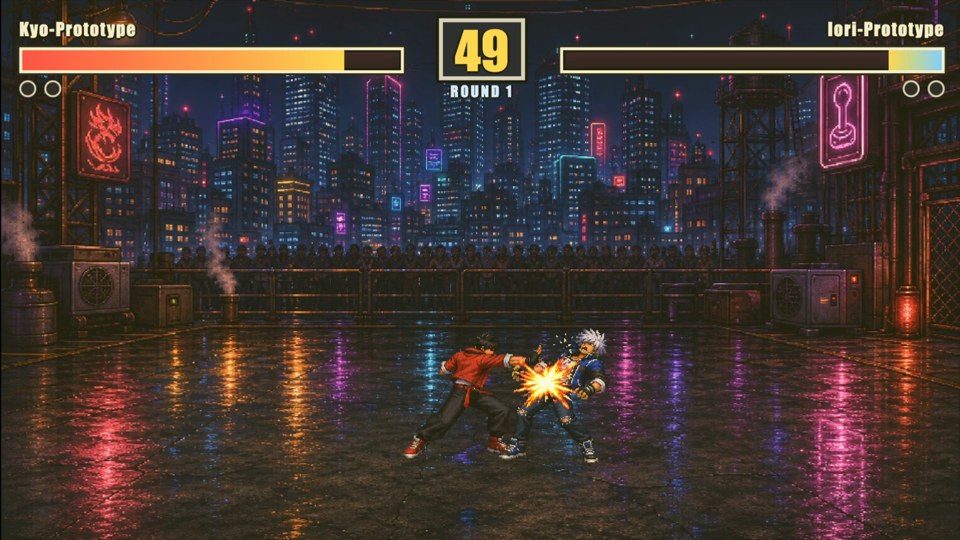}
  \caption{Two fighters trade blows with an impact spark on the neon stage.}
  \label{fig:app-kof}
\end{figure}

An arcade fighting game generated from a single-sentence prompt requesting a King of Fighters remake, with no reference image (Figure~\ref{fig:app-kof}). Two fighters trade combos under health bars, a round timer, and a neon-lit stage.

\subsection*{Dungeon Roguelike}
\begin{figure}[H]
  \centering
  \includegraphics[width=\linewidth]{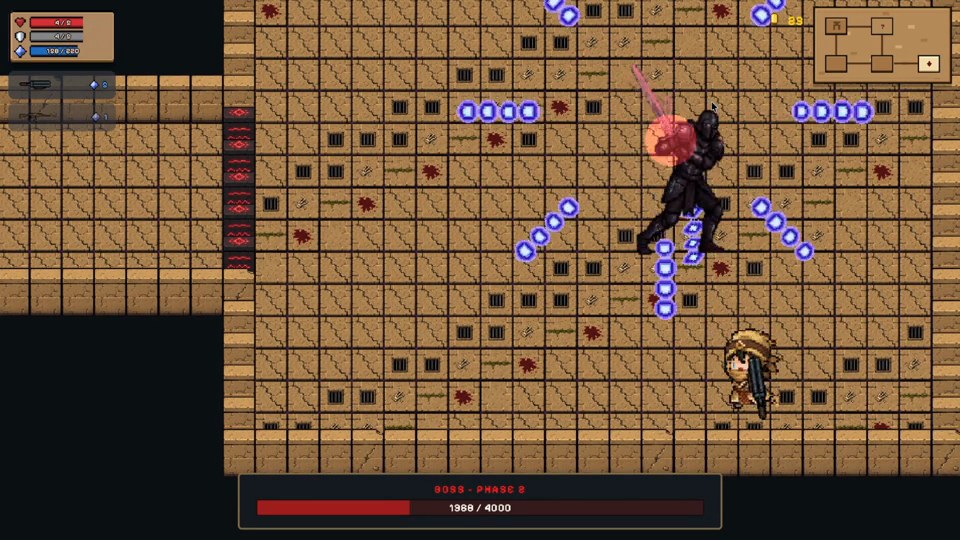}
  \caption{The player fires a spread of projectiles at an enemy inside a sealed combat room.}
  \label{fig:app-rog}
\end{figure}

A Soul Knight-style dungeon-shooter roguelike (Figure~\ref{fig:app-rog}). The map is built from rooms and corridors; entering a room seals the doors and starts combat. The player picks up weapons from chests, carries up to two at a time, and aims toward the mouse, with health, shield, and mana as stats.

\subsection*{Parkour Platformer}
\begin{figure}[H]
  \centering
  \includegraphics[width=\linewidth]{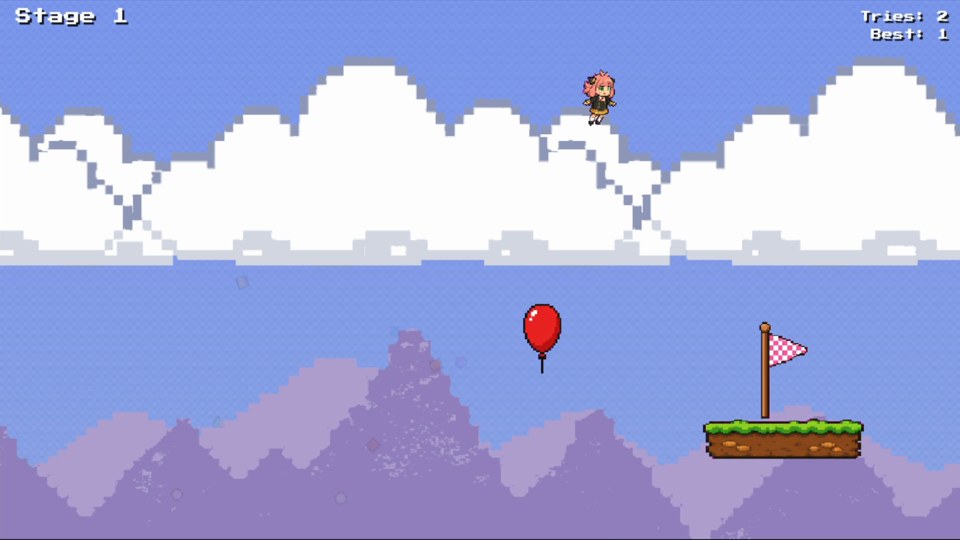}
  \caption{The protagonist floats above a balloon, with the goal flag on its platform.}
  \label{fig:app-parkour}
\end{figure}

A cartoon parkour platformer (Figure~\ref{fig:app-parkour}). The protagonist runs and dashes across a field of floating balloons; stepping on a balloon pops it and launches her upward, and the level is cleared by chaining precise balloon landings.

\subsection*{Civilization prototype}
\begin{figure}[H]
  \centering
  \includegraphics[width=\linewidth]{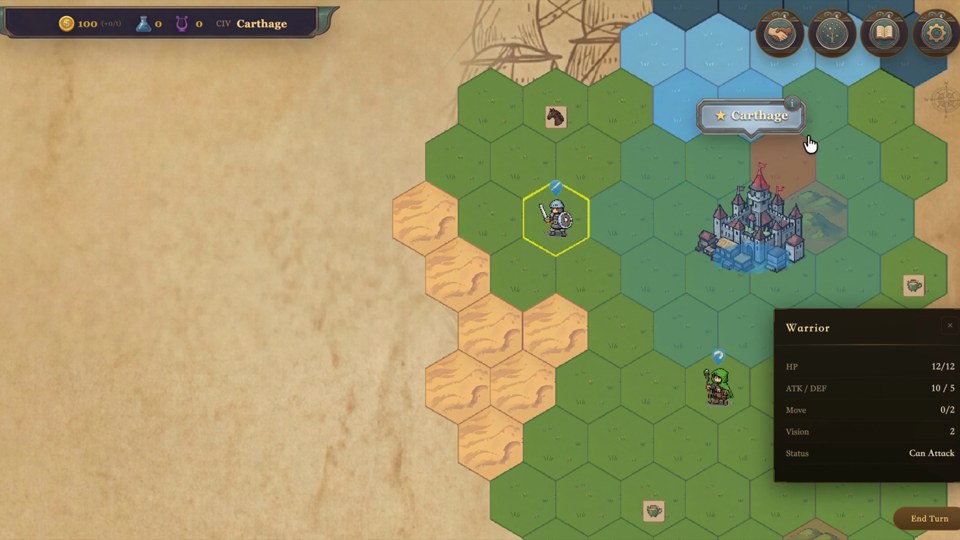}
  \caption{A selected warrior unit and its stats beside a city on the hex map.}
  \label{fig:app-civ}
\end{figure}

A simplified Civilization-style 4X strategy game on a hex map, rendered in a cartoon hand-drawn style with city management and unit combat (Figure~\ref{fig:app-civ}). It runs on a separated frontend and backend with multiplayer support.

\subsection*{AI Werewolf}
\begin{figure}[H]
  \centering
  \includegraphics[width=\linewidth]{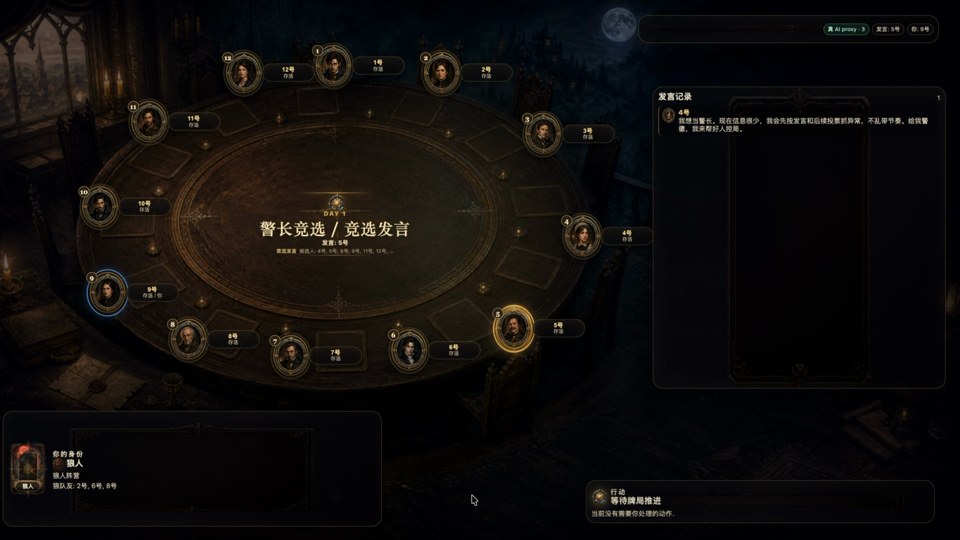}
  \caption{The round-table layout with AI player portraits and live speech text during the sheriff election.}
  \label{fig:app-werewolf}
\end{figure}

A single-player Werewolf game driven by the engine's built-in LLM module (Figure~\ref{fig:app-werewolf}). Eleven AI players fill a 12-role setup (villagers, werewolves, seer, hunter, witch, idiot), conducting sheriff elections, day discussions, and night actions; each AI's speech is rendered as on-screen text.

\subsection*{Hollow Knight to Demon Slayer}
\begin{figure}[H]
  \centering
  \includegraphics[width=\linewidth]{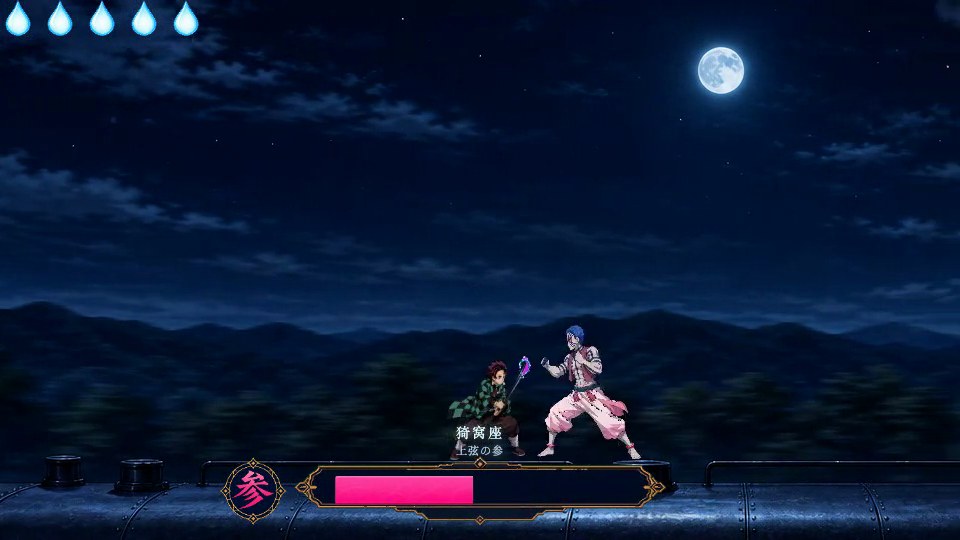}
  \caption{Tanjiro faces Upper Rank Three under the moon on the Mugen Train night field.}
  \label{fig:app-demonslayer}
\end{figure}

An IP transfer applied to the Hollow Knight-style boss-fight prototype from Appendix~\ref{sec:multimodal-input}: the system restyled the entire game into a Demon Slayer theme, recreating the Mugen Train battle between Tanjiro and Upper Rank Three while keeping the underlying combat logic unchanged (Figure~\ref{fig:app-demonslayer}).

\section{Adversarial Agent Team Case Study}
\label{sec:aat-case-study}

Three defects from one run of \vibegame with GPT-6.1-Sol on the GameCraft-Bench task Open-World Beast Tamer~\citep{gamecraftbench2026}, each reported by a different review role. Quotes are the agents' own reports.

\begin{figure}[H]
  \centering
  \includegraphics[width=\columnwidth]{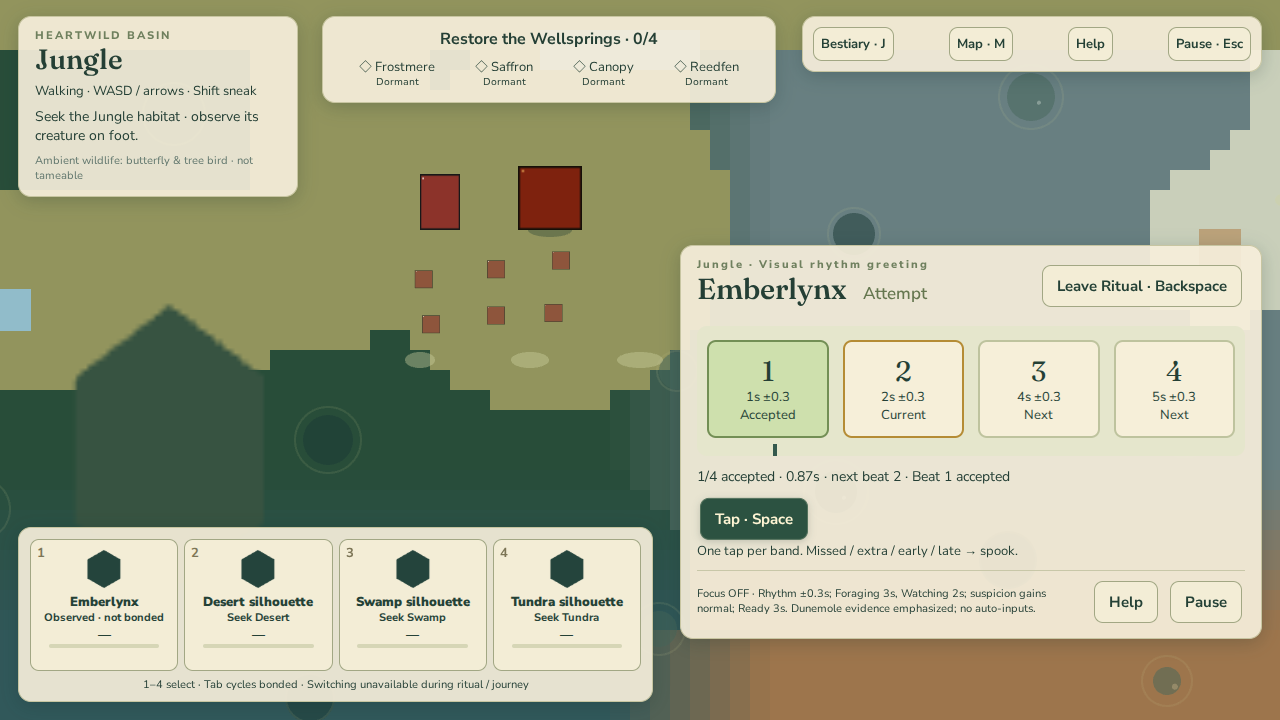}\\[2pt]
  \includegraphics[width=\columnwidth]{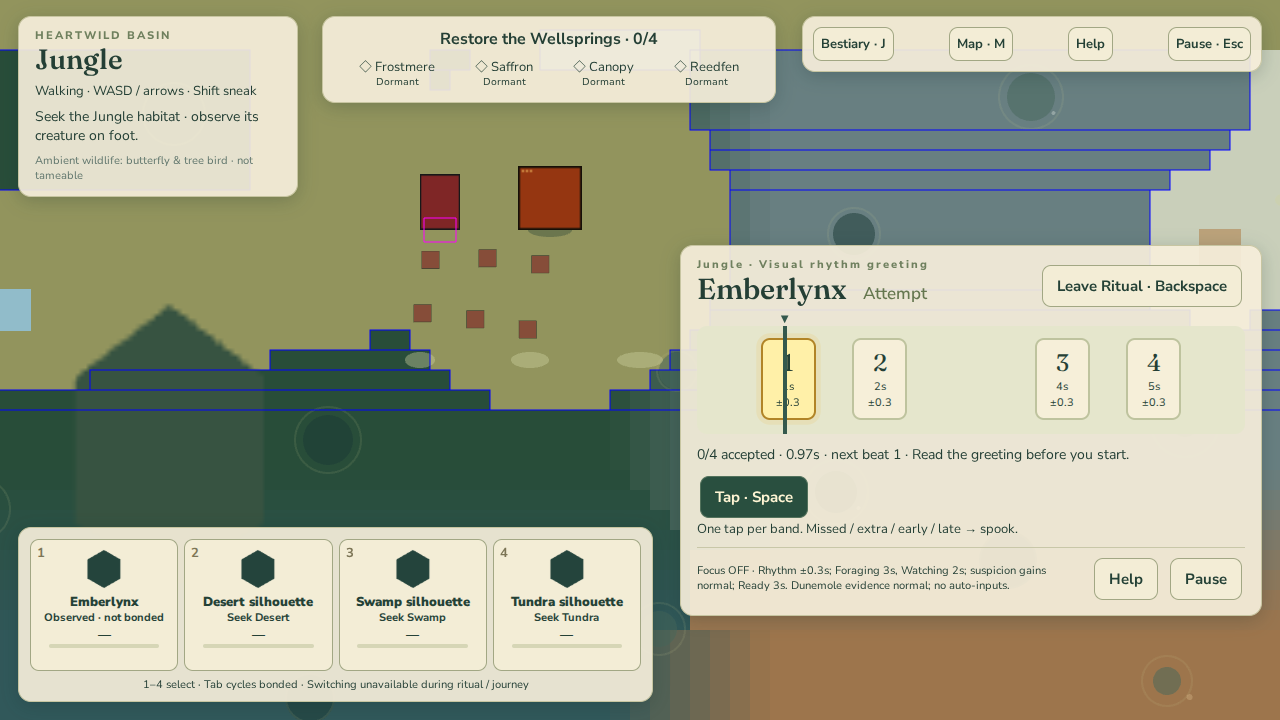}
  \caption{Auditor case, before (top) and after (bottom). After the repair, each band is placed and sized by its tapping window on one time axis. Placeholder art; the bottom image shows the physics debug overlay.}
  \label{fig:case-auditor}
\end{figure}

\paragraph{Auditor (static review).}
The PRD asks the player to tap ``when the timing indicator is inside its acceptance band'' (beats at 1, 2, 4, and 5\,s, $\pm$0.3\,s). Figure~\ref{fig:case-auditor} shows the game before and after the repair.
\begin{quote}\small
``AdventureUI.js:53--54 renders four equally spaced flex cards and an indicator at \texttt{elapsed / 6} \ldots{} Beat centers 1/2/4/5 and $\pm$0.3/$\pm$0.45 exist only as text; no band-position/width \ldots{} is computed from the scoring predicates. \ldots{} This is a missing timing-to-presentation mapping, not a visual-quality judgment.''
\end{quote}

\begin{figure}[H]
  \centering
  \includegraphics[width=\columnwidth]{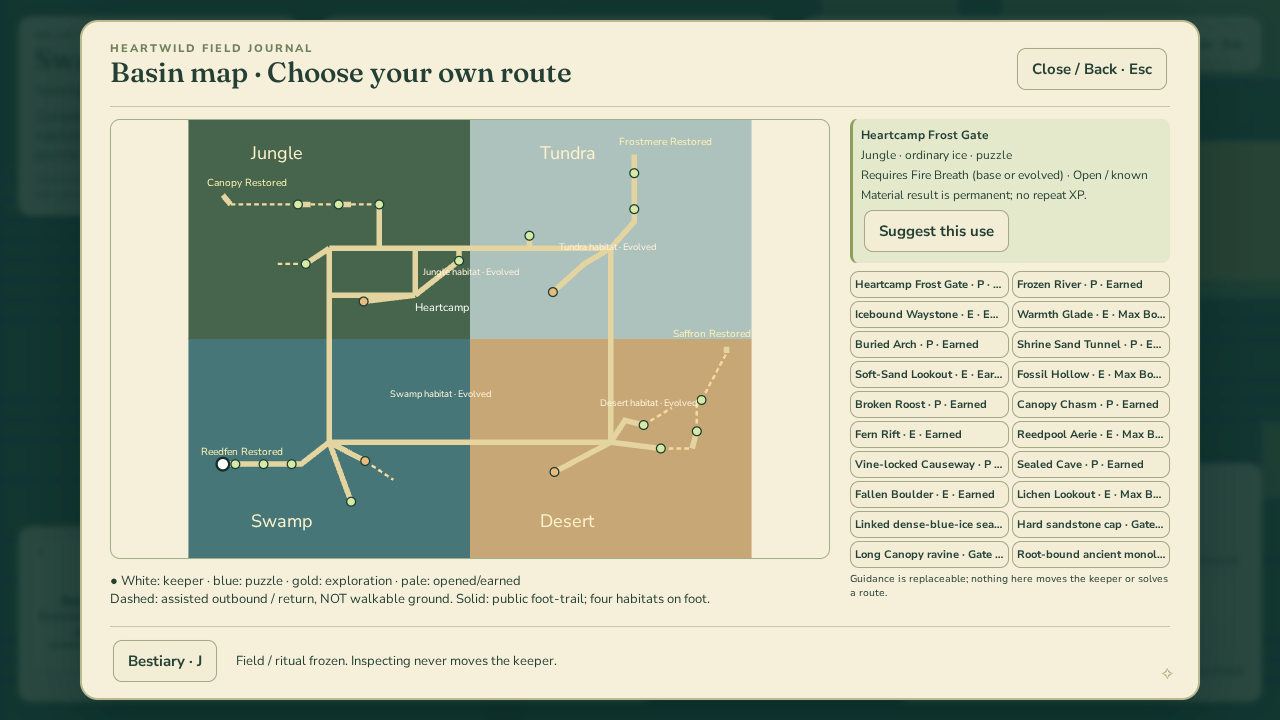}\\[2pt]
  \includegraphics[width=\columnwidth]{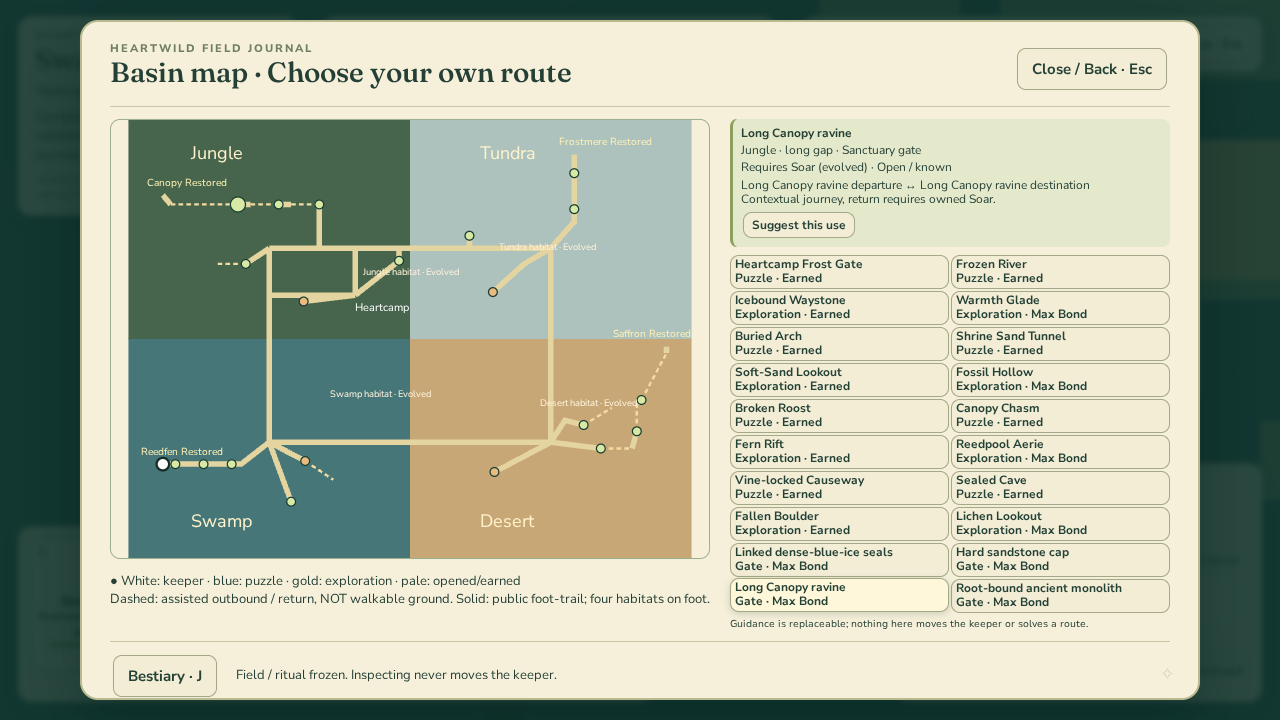}
  \caption{Player case, before (top) and after (bottom). After the repair, every entry shows its full name, type, and status.}
  \label{fig:case-player}
\end{figure}

\paragraph{Player (runtime check).}
An independent VLM pass first flagged the completed map; the player confirmed it and measured it in the running game. Figure~\ref{fig:case-player} shows the game before and after the repair.
\begin{quote}\small
``completed-map owned opportunity buttons visibly truncate names and type/status qualifiers with ellipses. Examples: Heartcamp Frost Gate $\cdot$ P\ldots{} \ldots{} label-overflow.json shows actual scroll widths exceeding client widths.''
\end{quote}

\begin{figure}[H]
  \centering
  \includegraphics[width=\columnwidth]{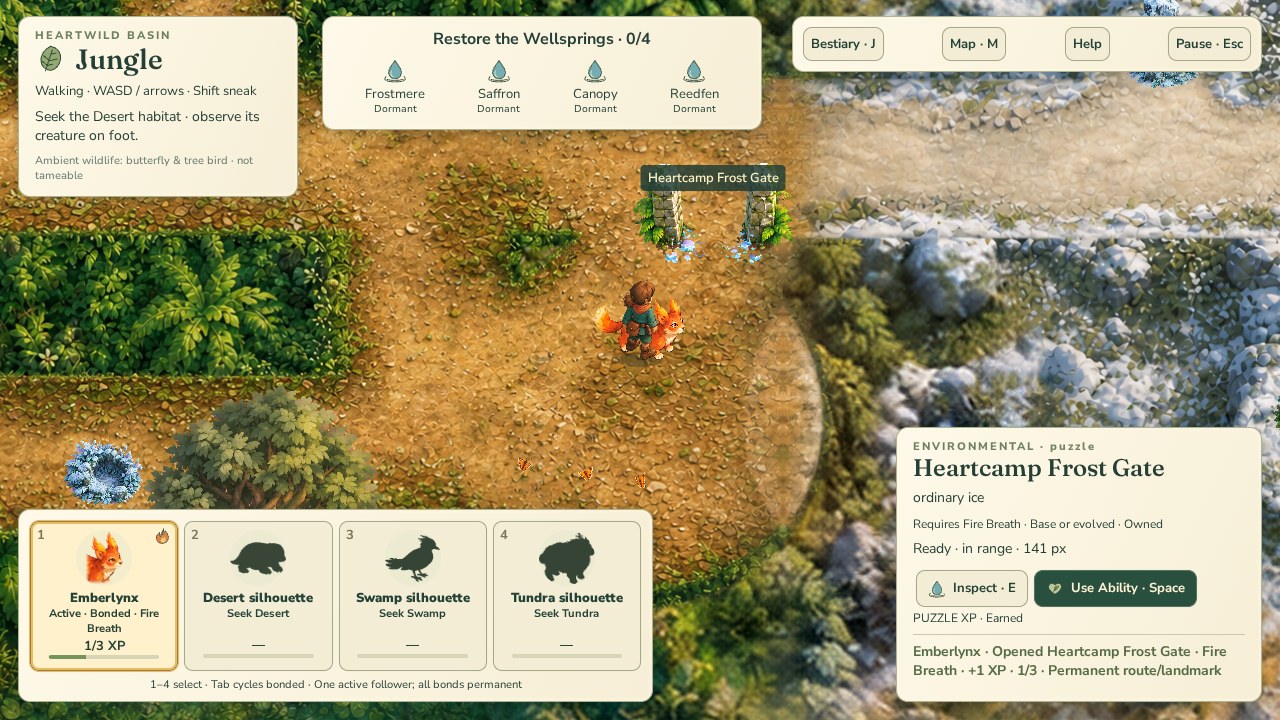}\\[2pt]
  \includegraphics[width=\columnwidth]{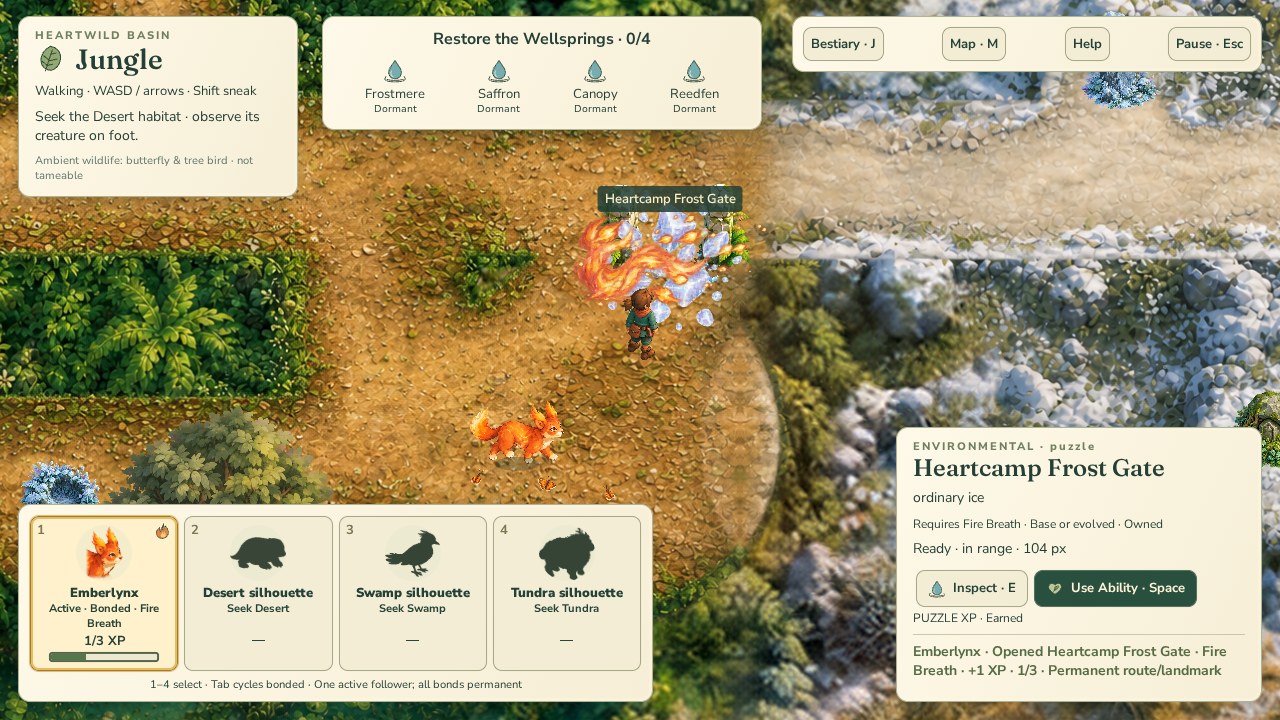}
  \caption{Reviewer case, before (top) and after (bottom). After the repair, the companion stands beside the keeper.}
  \label{fig:case-reviewer}
\end{figure}

\paragraph{Reviewer (final check).}
Found by stopping the game 120 frames after an ability and reading both characters' positions. Figure~\ref{fig:case-reviewer} shows the game before and after the repair.
\begin{quote}\small
``after a normal successful Fire Breath, waiting idle for two seconds leaves Emberlynx and the standing keeper at the exact same ground anchor, producing an inseparable/occluded central silhouette. VLM \ldots{} FAIL: the orange body reads as flames attached to the keeper, not a companion.''
\end{quote}

\section{Codex with the AI-Native Engine}
\label{sec:codex-engine}

To isolate the effect of the AI-native engine, we equip the native Codex harness with the engine of \vibegame.
Both settings use GPT-6.1-Sol on all 30 tasks in the PA2G setting with the same asset library.
With the engine, the overall score of Codex rises from 64.44 to 67.13, with gains in all four categories (Table~\ref{tab:codex-engine}).
Since the art in both settings comes from the same asset library, we attribute the improvement to the engine; the largest gain, in Content Depth, suggests that the structured project representation helps agents expand game content.

\begin{table}[H]
  \centering
  \caption{Codex with and without the AI-native engine of \vibegame (GPT-6.1-Sol, PA2G, all 30 tasks). Columns as in Table~\ref{tab:main}.}
  \setlength{\tabcolsep}{10pt}
  \begin{tabular}{@{}lccccc@{}}
    \toprule
    \textbf{Harness} & \textbf{M} & \textbf{D} & \textbf{V} & \textbf{A} & \textbf{Overall} \\
    \midrule
    Codex     & 80.21 & 58.42 & 57.51 & 66.52 & 64.44 \\
    w/ Engine & 82.41 & 62.37 & 58.94 & 68.75 & 67.13 \\
    \bottomrule
  \end{tabular}
  \label{tab:codex-engine}
\end{table}

\section{Limitation of Self-Evolution}
\label{sec:evolution-limitation}

We observe that when \vibegame continuously develops games of different types, the experience accumulated by self-evolution can instead lower agent performance.
Future work should analyze this phenomenon further and separate general experience from genre-specific experience, so that each can be injected adaptively.
This observation is also one of the motivations for our experimental design: the main experiments use \vibegame without any accumulated experience, and we evaluate self-evolution separately in the ablation study.

\end{document}